\documentclass[aps,prd,groupedaddress,nofootinbib,amssymb,eqsecnum,epsfig]{revtex4}
\usepackage{graphicx}
\usepackage{bm}
\usepackage{amsmath}
\usepackage{color}
\usepackage{amsfonts}

\begin{document}
\newcommand{\newc}{\newcommand}

\newcommand{\ben}{\begin{eqnarray}}
\newcommand{\een}{\end{eqnarray}}
\newc{\be}{\begin{equation}}
\newc{\ee}{\end{equation}}
\newc{\ba}{\begin{eqnarray}}
\newc{\ea}{\end{eqnarray}}
\newc{\bea}{\begin{eqnarray*}}
\newc{\eea}{\end{eqnarray*}}
\newc{\D}{\partial}
\newc{\ie}{{\it i.e.} }
\newc{\eg}{{\it e.g.} }
\newc{\etc}{{\it etc.} }
\newc{\etal}{{\it et al.}}
\newcommand{\nn}{\nonumber}
\newc{\ra}{\rightarrow}
\newc{\lra}{\leftrightarrow}
\newc{\lsim}{\buildrel{<}\over{\sim}}
\newc{\gsim}{\buildrel{>}\over{\sim}}
\newc{\cU}{{\cal U}}
\newc{\cE}{{\cal E}}
\newc{\rd}{{\rm d}}
\newc{\bN}{{\bar N}}
\newc{\cH}{{\cal H}}

\title{Dark energy survivals in massive gravity after GW170817: SO(3) invariant}

\author{
Lavinia Heisenberg$^{1}$ and 
Shinji Tsujikawa$^{2}$ }

\affiliation{
$^1$Institute for Theoretical Studies, ETH Zurich, Clausiusstrasse 47, 8092 Zurich, Switzerland\\
$^2$Department of Physics, Faculty of Science, Tokyo University of Science, 1-3, Kagurazaka,
Shinjuku-ku, Tokyo 162-8601, Japan}

\date{\today}

\begin{abstract}

The recent detection of the gravitational wave signal GW170817 together with an electromagnetic counterpart GRB 170817A from the merger of two neutron stars 
puts a stringent bound on the tensor propagation speed. 
This constraint can be automatically satisfied in the framework 
of massive gravity. In this work we consider a 
general $SO(3)$-invariant massive gravity with five 
propagating degrees of freedom and derive the conditions for 
the absence of ghosts and Laplacian instabilities 
in the presence of a matter perfect fluid on the flat 
Friedmann-Lema\^{i}tre-Robertson-Walker (FLRW) 
cosmological background. 
The graviton potential containing the 
dependence of three-dimensional metrics and a 
fiducial metric coupled to a temporal scalar field
gives rise to a scenario of the late-time cosmic 
acceleration in which the dark energy equation of state $w_{\rm DE}$ is equivalent to $-1$ or varies in time.
We find that the deviation from the value $w_{\rm DE}=-1$
provides important contributions to the quantities 
associated with the stability conditions of tensor, vector, 
and scalar perturbations. 
In concrete models, 
we study the dynamics of dark energy arising from 
the graviton potential and show that there exist
viable parameter spaces in which neither ghosts 
nor Laplacian instabilities are present 
for both $w_{\rm DE}>-1$ 
and $w_{\rm DE}<-1$. 
We also generally obtain the effective gravitational coupling 
$G_{\rm eff}$ with non-relativistic matter as well as the gravitational 
slip parameter $\eta_s$ associated with the observations of large-scale 
structures and weak lensing. 
We show that, apart from a specific case, the two quantities 
$G_{\rm eff}$ and $\eta_s$ are similar to those in general relativity 
for scalar perturbations deep inside the sound horizon.

\end{abstract}


\maketitle

\section{Introduction}

The large-distance modification of gravity has been under 
active study over the past two 
decades \cite{rev1,rev2,rev3,rev4,rev5,rev6,rev7,rev8}. 
This is mostly attributed to the constantly accumulating observational evidence of the late-time cosmic 
acceleration \cite{SN1,SN2,SN3,WMAP,Planck,BAO}. 
In General Relativity (GR), we need to introduce an unknown matter component dubbed dark energy 
to account for this phenomenon.
In modified gravitational theories, new degrees of freedom (DOFs) arising from the breaking of gauge symmetry in GR 
can be the source for the acceleration.

The new DOFs appearing in modified gravitational theories 
usually possess the properties of scalar or vector fields 
besides two tensor polarizations. 
The most general second-order scalar-tensor theories 
with one scalar and two tensor DOFs are known 
as Horndeski theories \cite{Horndeski}. 
The application of theories within the 
framework of Horndeski theories--such as $f(R)$ 
gravity \cite{fR1,fR2} 
and Galileons \cite{Ga1,Ga2}-- to the late-time 
cosmic acceleration has been extensively performed 
in the literature \cite{fRcos1,fRcos2,fRcos3,fRcos4,Gacos1,Gacos2}. 
In Refs.~\cite{Heisenberg,Tasinato,Allys,Jimenez} 
the authors constructed 
second-order massive vector-tensor theories 
with a vector field $A^{\mu}$ coupled to gravity, 
which propagate five DOFs (one scalar, 
two vectors, two tensors). 
Such theories can also lead to the 
cosmic acceleration \cite{Procacosmo,Procacosmo2}, 
while satisfying local gravity 
constraints in the solar system \cite{screening1,screening2}.

Massive gravity is also the way of modifying gravity at large 
distances by giving a tiny mass to 
the graviton \cite{mareview}. 
In general there are six propagating DOFs in 
massive gravity, one of which behaves as a ghost. 
Fierz and Pauli (FP) \cite{FP} constructed a Lorentz-invariant linear theory of massive gravity without the ghost on the Minkowski background. However, the FP theory is plagued by a problem 
of the so-called van Dam-Veltman-Zakharov (vDVZ) 
discontinuity with which the massless limit does not 
recover GR \cite{dis1,dis2}. 
Vainshtein \cite{Vainshtein} argued that the 
vDVZ discontinuity originates from the breakdown of 
linear FP theory in the massless limit and that nonlinearities can cure the problem. However, the nonlinear extension of FP theory generally suffers from the appearance of a so-called 
Boulware-Deser (BD) ghost \cite{BD}.

In 2010, de Rham, Gabadadze, and Tolley (dRGT) \cite{dRGT} constructed a unique nonlinear theory of 
Lorentz-invariant massive gravity without 
the BD ghost (see also Refs.~\cite{deRham10,Hassan}).
Applying the dRGT theory to cosmology, it was first recognized that there are no viable expanding solutions 
on the flat FLRW background \cite{DAmico} (see also \cite{deRham:2010tw,deRham:2011by,Burrage:2011cr,Chamseddine:2011bu,Koyama:2011xz,Gratia:2012wt,Comelli:2011zm,Comelli:2012db,Volkov:2012zb}). 
If we replace the Minkowski fiducial metric in the 
original dRGT theory with a de Sitter or FLRW fiducial 
metric, this allows for the existence of self-accelerating 
or non self-accelerating solutions \cite{Emir,Tolley,Naruko}.
The resulting self-accelerating solution can be relevant to the 
late-time cosmic acceleration, but it suffers from 
the strong coupling problem where two kinetic terms 
among five propagating DOFs exactly vanish \cite{Emir}. 
This is associated with the nonlinear ghost instability,  
which manifests itself on the anisotropic cosmological 
background \cite{DGM}. Apart from the case in which 
matter species have a specific coupling to 
the metric \cite{maco1,maco2,Heisenberg:2014rka,Gumrukcuoglu:2014xba,Heisenberg:2016spl},  
the cosmological solutions in dRGT theory are generally 
plagued by the problem of ghost instabilities (including 
the appearance of a Higuchi ghost \cite{Higuchi} 
for non self-accelerating solutions). 

If we break the Lorentz-invariance, it is possible to avoid 
the aforementioned problems of Lorentz-invariant massive 
gravity theories. Indeed, the extension of the FP action 
to a Lorentz-violating form does not give to the vDVZ 
discontinuity \cite{Rubakov}. 
The expansion around the Minkowski background shows 
that, for the graviton mass $m$, the low-energy effective theory 
is valid up to the strong coupling scale 
$\sqrt{m M_{\rm pl}}$, 
where $M_{\rm pl}=2.435 \times 10^{18}$~GeV is the 
reduced Planck mass. The construction of $SO(3)$-invariant 
theories of massive gravity was performed in Ref.~\cite{Dub1} 
by imposing the residual reparametrization 
symmetry $x^i \to x^i+\xi^i(t)$, where $x^i$ are 
spatial coordinates and $\xi^i$ is a 
time-dependent shift. The application of such 
rotational invariant massive gravity to cosmology 
was carried out in Refs.~\cite{Dub2,Dub3,Beb,Blas,Gaba} 
(see also Refs.~\cite{Lin1,Lin2,Lin3}).
The resulting solutions can lead to the late-time cosmic acceleration \cite{Dub3} without theoretical 
pathologies.

In $SO(3)$-invariant massive gravity theories, 
Comelli {\it et al.} \cite{Comelli1,Comelli2} 
derived a most general form of the graviton potential with five propagating DOFs. {}From the Hamiltonian analysis 
the extra ghost DOF among six DOFs is eliminated at fully nonperturbative level.
Applying such general massive gravity theories to cosmology, the background dynamics is parametrized by a time-dependent function ${\cal U}$. This can lead to the cosmic acceleration 
with a dynamically varying dark energy equation of state $w_{\rm DE}$ \cite{Comelli3}.
If we consider a fiducial metric coupled to a temporal 
scalar field $\phi$ in the graviton potential, 
there exists a self-accelerating solution 
($w_{\rm DE}=-1$) free from the strong coupling problem 
associated with vanishing kinetic terms \cite{Lan}.

So far, the stability analysis against perturbations 
on the FLRW background in general $SO(3)$-invariant massive gravity theories has been restricted to the case 
without matter. In this paper we take into 
account a matter perfect fluid as a form of the 
Schutz-Sorkin action \cite{Sorkin}, which has an 
advantage for dealing with vector perturbations 
appropriately \cite{DGS,Procacosmo}.
We will generally obtain conditions for the absence of ghosts and Laplacian instabilities of tensor, vector, and scalar perturbations and apply them to a concrete model of 
Lorentz-violating massive gravity with the 
late-time cosmic acceleration. As in the model discussed in 
Ref.~\cite{Lan} we allow for the existence of a 
non-trivial fiducial metric multiplied by a temporal scalar, 
but we will study more 
general cases in which the dark energy equation of 
state varies in the region $w_{\rm DE}>-1$ 
or $w_{\rm DE}<-1$. For the purpose of constraining 
$SO(3)$-invariant massive gravity from observations 
of large-scale structures and weak lensing, 
we will also derive the effective gravitational coupling 
with nonrelativistic matter and the gravitational 
slip parameter.

We note that the recent detection of gravitational waves from 
a binary neutron star merger placed the bound 
$-3 \times 10^{-15} \le c_T/c-1 \le 7 \times 10^{-16}$, where 
$c_T$ is the tensor propagation speed and $c$ is 
the speed of light \cite{LIGO}. 
In our $SO(3)$-invariant massive gravity the tensor speed $c_T$ is equivalent to $c$, so it evades the 
tight bound constrained  from GW170817. 
For the graviton mass the observations of GW170104 
put the upper limit 
$m \le 7.7 \times 10^{-23}$~eV \cite{LIGO2}.
The massive gravity relevant to the late-time 
cosmic acceleration has the graviton mass of 
order $m \lesssim 10^{-33}$~eV, which
is well consistent with the GW170104 bound.
The Lorentz-violating dark energy scenario with
such a tiny graviton mass does not contradict with 
observations associated with the 
Lorentz violation either. The same is true for the doubly coupled massive gravity scenario \cite{Gumrukcuoglu:2014xba,Heisenberg:2016spl}.

This paper is organized as follows. 
In Sec.\,\ref{theorysec} we briefly review general 
Lorentz-violating massive gravity with the $SO(3)$ invariance.
In Sec.\,\ref{backsec} we discuss the cosmology 
on the flat FLRW background by paying particular attention to the dynamics of 
dark energy arising from the graviton potential.
In Sec.\,\ref{masssec} we revisit how the Lorentz-violating 
mass terms arise after expanding the action up to second 
order in perturbations. 
In Sec.\,\ref{tenvecsec} we derive conditions for avoiding  ghosts and Laplacian instabilities of tensor and vector 
perturbations in the presence of matter. 
In Sec.\,\ref{scasec} we obtain stability conditions of 
scalar perturbations and identify
parameter spaces consistent with all the stability conditions 
for a concrete model. 
In Sec.\,\ref{growthsec} we study the growth of 
non-relativistic matter perturbations and derive the 
effective gravitational coupling and the gravitational 
slip parameter under a quasi-static approximation 
deep inside the sound horizon.
Sec.\,\ref{concludesec} is devoted to conclusions.

\section{General $SO(3)$-invariant massive gravity}
\label{theorysec}

The Lorentz-violating massive gravity with three-dimensional Euclidean symmetry are characterized by the four-dimensional metric tensor $g_{\mu \nu}$ and scalar fields 
$(\phi, \phi^i)$, where $i=1,2,3$. 
The theory respects the three-dimensional 
internal symmetry
\be
\phi^i \to \Lambda^{i}_{j} \phi^j\,,\qquad 
\phi^i \to \phi^i+C^i\,,
\ee
where $\Lambda^{i}_{j}$ is a $SO(3)$ rotational operator, 
and $C^i$ are constants. 
Due to this symmetry, the fields $\phi^i$ appear in the action 
only as a form of the fiducial metric 
$e_{\mu \nu}=\delta_{ij} \partial_{\mu}\phi^i
\partial_{\mu}\phi^j$.
Then, the massive graviton potential $V$ generally depends 
on $g_{\mu \nu}, e_{\mu \nu}, \phi, \partial_{\mu} \phi$. 
The higher-order derivative terms are not taken into account
to avoid the Ostrogradski instability.

We introduce the ten scalar quantities
\be
{\cal N} \equiv \frac{1}{\sqrt{-g^{\mu \nu} \partial_{\mu} \phi \partial_{\nu} \phi}}\,,\qquad
{\cal N}^i \equiv {\cal N}n^{\mu} \partial_{\mu} \phi^i\,,
\qquad 
\Gamma^{ij} \equiv \left( g^{\mu \nu}+n^{\mu}n^{\nu} 
\right) \partial_{\mu}\phi^i \partial_{\nu}\phi^j\,, 
\label{scalarqu}
\ee
where $n^{\mu}={\cal N}g^{\mu \nu}\partial_{\nu}\phi$
is a unit vector, and $\Gamma^{ij}=\Gamma^{ji}$. 
In the following, we choose the unitary gauge in which 
the scalar components $\phi$ and $\phi^i$ are identical to 
the time $t$ and the spatial vector $x^i$, respectively, 
\be
\phi=t\,,\qquad \phi^i=x^i\,.
\label{unitary}
\ee
For this gauge choice the quantities in Eq.~(\ref{scalarqu}) 
reduce to ${\cal N}=1/\sqrt{-g^{00}}$, 
${\cal N}^i={\cal N}n^i$,  
and $\Gamma^{ij}=g^{ij}+n^in^j$, where 
$n^i={\cal N}g^{i0}$. 
In this case, the quantities ${\cal N}$, ${\cal N}^i$, and 
$\Gamma^{ij}$ correspond to the lapse $N$, the shift 
vector $N^i$, and the three-dimensional metric 
$\gamma^{ij}$ in the ADM formalism \cite{ADM}, respectively,  
with the line element 
\be
ds^2=g_{\mu \nu}dx^{\mu}dx^{\nu}
=-N^2 dt^2+\gamma_{ij} \left( dx^i+N^i dt \right)
\left( dx^j+N^j dt \right)\,.
\label{ADM}
\ee

The most general massive graviton potential $V$ with five 
propagating DOFs is parametrized by two 
$SO(3)$-invariant functions 
${\cal U}$ and ${\cal E}$ \cite{Comelli1,Comelli2,Comelli3}.
The function ${\cal U}$ depends on the special combination 
\be
{\cal K}^{ij}=\gamma^{ij}-\xi^i \xi^j\,,
\ee
where the new shift variables $\xi^i$ are related to $N^i$
according to
\be
N^i-N\xi^i=-{\cal U}_{ij}^{-1}{\cal E}_j\,.
\label{NiN}
\ee
Here, ${\cal U}_{ij}^{-1}$ is the inverse of the matrix 
${\cal U}_{ij} \equiv \partial^2 {\cal U}/\partial \xi^i 
\partial \xi^j$, ${\cal E}$ is a function containing the 
dependence of $\gamma^{ij}$ and $\xi^i$, and 
${\cal E}_j \equiv \partial {\cal E}/
\partial \xi^j$. 
If $\cE_j=0$, then the vector $\xi^i$ is directly related to 
the shift $N^i$ as $\xi^i=N^i/N$. 
In this case, the quantity ${\cal K}^{ij}$ reduces to 
the three-dimensional metric  
$g^{ij}=\gamma^{ij}-N^i N^j/N^2$.

The massive graviton potential that propagates 
five DOFs is restricted to the form \cite{Comelli1,Comelli2,Comelli3} 
\be
V={\cal U}+\frac{1}{N} \left( {\cal E} 
-{\cal U}_i\,{\cal U}_{ij}^{-1} {\cal E}_j \right)\,,
\label{Vdef}
\ee
where ${\cal U}_i \equiv \partial {\cal U}/\partial \xi^i$.
In the unitary gauge the functions ${\cal U}$ and ${\cal E}$ 
also depend on the fiducial metric $e_{ij}=\delta_{ij}$ 
and the temporal scalar $\phi=t$, 
such that\footnote{This formulation does not accommodate 
a Lorentz-violating minimal theory of massive gravity proposed in Refs.~\cite{mini1,mini2}, as the latter contains the 
time derivative of $\gamma_{ij}$. 
The non-local theory of massive gravity \cite{nl1,nl2,nl3} 
is also beyond our framework.} 
\be
{\cal U}={\cal U}({\cal K}^{ij}, \delta_{ij}, \phi)\,,
\qquad 
{\cal E}={\cal E} (\gamma^{ij}, \xi^i, \delta_{ij}, \phi)\,.
\label{UE0}
\ee
We will consider the combination of 
$\delta_{ij}$ and $\phi$ given by \cite{Lan}
\be
f_{ij}=b^2(\phi) \delta_{ij}\,,
\ee
where $b(\phi)$ is an arbitrary function of $\phi$. 
In this case, the $SO(3)$-invariant functions 
${\cal U}$ and ${\cal E}$ have the dependence
\be
{\cal U}={\cal U}({\cal K}^{ij}, f_{ij})\,,
\qquad 
{\cal E}={\cal E} (\gamma^{ij}, \xi^i, f_{ij})\,.
\label{UE}
\ee
The massive gravity theory invariant under the shift symmetry 
$\phi \to \phi+c$ corresponds to $b(\phi)=1$.
For the choice $b(\phi)=e^{M\phi}$, the above prescription 
accommodates the theory invariant under the diatonic
global symmetry $\phi \to \phi+c$ and 
$\phi^i \to e^{-Mc} \phi^i$ \cite{Lan}.

Besides the massive graviton potential, we also take into 
account the Ricci scalar $R$ in the action and the matter field 
$\Psi_M$ minimally coupled to gravity.
Namely, we focus on the massive gravity theories 
with the action 
\be
\label{Lag}
S=S_{\rm EH}+S_{\rm mg}
+S_M\,,
\ee
where the Einstein-Hilbert term $S_{\rm EH}$ and  
the massive graviton term $S_{\rm mg}$ are given,  
respectively, by
\ba
S_{\rm EH}
&=& M_{\rm pl}^2 \int d^4 x \sqrt{-g}\,\frac{R}{2}\,,\label{SEH}\\
S_{\rm mg}&=&-M_{\rm pl}^2 \int d^4 x \sqrt{-g}\,m^2\,V\,.
\label{Smg}
\ea
Here, $g$ is the determinant of $g_{\mu \nu}$, $m$ 
is the graviton mass, and $V$ is the graviton 
potential of the form (\ref{Vdef}).

To describe a perfect fluid for the matter field $\Psi_M$, 
we consider the Schutz-Sorkin action \cite{Sorkin,DGS}
\be
S_M=-\int d^{4}x \left[ \sqrt{-g}\,\rho_M(n)
+J^{\mu}(\partial_{\mu}\ell+\mathcal{A}_1
\partial_{\mu}\mathcal{B}_1+\mathcal{A}_2
\partial_{\mu}\mathcal{B}_2) \right]\,,
\label{SM}
\ee
where $\rho_M$ is the fluid density that depends on its 
number density defined by 
\be
n=\sqrt{\frac{J^{\alpha}J^{\beta}g_{\alpha\beta}}{g}}\,,
\label{num}
\ee
with $J^{\mu}$ being a vector field of weight one. 
The quantity $\ell$ corresponds to the scalar mode, 
whereas $\mathcal{A}_1, \mathcal{A}_2, 
\mathcal{B}_1, \mathcal{B}_2$ are scalar quantities 
whose perturbations are associated with the vector modes.
In massive gravity theories the vector modes generally propagate, so it is convenient to use the Schutz-Sorkin 
action to describe the propagation of vector modes 
in the matter sector (as in the case of vector-tensor 
theories \cite{Procacosmo,Procacosmo2}).
 
\section{FLRW background and dark energy dynamics}
\label{backsec}

For the massive gravity theory given by the action (\ref{Lag}) the equations of motion on the flat FLRW background 
were derived in Ref.~\cite{Comelli3}.
In Ref.~\cite{Lan} the authors proposed a concrete 
$SO(3)$-invariant massive gravity model with a 
nontrivial fiducial metric to discuss the property of 
de Sitter solutions. Here we will revisit them in the presence 
of matter and study the background dynamics relevant to 
the late-time cosmic acceleration. 
Compared to Ref.~\cite{Lan}, our analysis is more general 
in that the expansion rate arising from $b(t)$ in the metric 
$f_{ij}$ is not necessarily equivalent to the Hubble 
expansion rate.

\subsection{Background equations}

Let us consider the flat FLRW spacetime described 
by the line element 
\be
\overline{ds}^2=-\bar{N}^2(t)dt^2
+a^2(t)\delta_{ij}dx^idx^j\,,
\label{backmet}
\ee
where $\bar{N}(t)$ is the background value of the lapse 
$N$, and $a(t)$ is the time-dependent scale factor. 
Here and in the following, we use an overbar to 
represent background quantities.
Since $\bar{N}^i=\bar{\xi}^i=
\bar{{\cal U}}_{ij}^{-1}\bar{{\cal E}}_j=0$ 
on the FLRW background, 
the massive graviton potential (\ref{Vdef}) 
reduces to $\bar{V}=\bar{\cal U}+\bar{{\cal E}}/\bar{N}$.
Up to boundary terms, the sum of the two actions 
(\ref{SEH}) and (\ref{Smg}) reads
\be
\bar{S}_{\rm EH}+\bar{S}_{\rm mg}
=-M_{\rm pl}^2 \int d^4 x
\left[ \frac{3a \dot{a}^2}{\bar{N}}
+a^3 m^2 \left( {\bar N} \bar{\cU}+\bar{\cE} \right) 
\right]\,,
\ee
where an overdot represents a derivative with respect to $t$. 
The two functions in Eq.~(\ref{UE}) have the dependence 
$\bar{{\cal U}}=\bar{{\cal U}}(\bar{\gamma}^{ij}, f_{ij})$ and 
$\bar{{\cal E}}=\bar{{\cal E}}(\bar{\gamma}^{ij}, f_{ij})$, where 
$\bar{\gamma}^{ij}=a^{-2}(t)\delta^{ij}$ and 
$f_{ij}=b^{2}(t)\delta_{ij}$.

On the FLRW background (\ref{backmet}) the matter action 
(\ref{SM}) reduces to 
$\bar{S}_M=-\int d^4x [\sqrt{-\bar{g}}\,\bar{\rho}_M
+\bar{J}^{\mu} \partial_{\mu} \bar{\ell}]$.
The temporal component $\bar{J}^0$ corresponds to the total fluid number ${\cal N}_0$, which is constant.
From Eq.~(\ref{num}) the number density $n_0$ 
is given by 
\be
n_0=\frac{{\cal N}_0}{a^3}\,, 
\ee
and hence $\bar{J}^0=n_0 a^3$. 
Then, the matter action (\ref{SM}) is expressed as  
\be
\bar{S}_M=-\int d^4x \,\left( \bar{N}a^3 \bar{\rho}_M
+n_0 a^3 \dot{\bar{\ell}} \right)\,.
\label{PM} 
\ee
For the variation of the total action 
$\bar{S}=\bar{S}_{\rm EH}+\bar{S}_{\rm mg}+\bar{S}_M$, 
we exploit the fact that 
$\partial \bar{\cU}/\partial a=-6\,{\rm d}\cU/a$ and 
$\partial \bar{\cE}/\partial a=-6\,{\rm d}\cE/a$, where 
the derivatives ${\rm d}\cU$ and ${\rm d}\cE$ are 
defined by 
\be
\overline{\frac{\partial \cU}{\partial {\cal K}^{ij}}}
\equiv {\rm d}\cU \bar{\gamma}_{ij}\,,\qquad
\overline{\frac{\partial \cE}{\partial {\gamma}^{ij}}} 
\equiv {\rm d}\cE \bar{\gamma}_{ij}\,.
\label{dUdef}
\ee
Varying the action $\bar{S}$ with respect to $\bar{N}$ and 
$a$, respectively, we obtain the background equations of motion 
\ba
& &
3M_{\rm pl}^2H^2=\rho_{\rm mg}+\bar{\rho}_M\,,
\label{back1}\\
& &
M_{\rm pl}^2 \left( \frac{2\dot{H}}{\bar{N}}
+3H^2 \right)
=-P_{\rm mg}-\bar{P}_M\,,
\label{back2}
\ea
where 
\be
H \equiv \frac{\dot{a}}{\bar{N}a}
\ee
is the Hubble expansion rate, and $\bar{P}_M$ is 
the matter pressure defined by 
\be
\bar{P}_M \equiv -n_0 \frac{\dot{\bar{\ell}}}{\bar{N}}
-\bar{\rho}_M\,.
\label{PMva}
\ee
The quantities $\rho_{\rm mg}$ and $P_{\rm mg}$, which 
correspond to the energy density and the pressure 
arising from the graviton potential respectively, are 
given by 
\be
\rho_{\rm mg}=M_{\rm pl}^2 m^2 \bar{\cU}\,,\qquad
P_{\rm mg}=M_{\rm pl}^2 m^2 
\left[ 2{\rm d} \cU-\bar{\cU}
+\frac{1}{\bar{N}} \left( 2{\rm d}\cE-\bar{\cE} \right)
\right]\,.
\label{rhoPmg}
\ee

The matter sector obeys the continuity equation 
\be
\dot{\bar{\rho}}_M
+3\bar{N}H \left( \bar{\rho}_M+\bar{P}_M
\right)=0\,.
\label{conm}
\ee
On using Eqs.~(\ref{back1}) and (\ref{back2}) with 
Eq.~(\ref{conm}), the massive gravity sector also satisfies 
$\dot{\rho}_{\rm mg}+3\bar{N}H \left( \rho_{\rm mg}
+P_{\rm mg} \right)=0$. This translates to 
\be
\dot{\bar{\cU}}+6\bar{N}H{\rm d}\cU
+3H \left( 2{\rm d} \cE-\bar{\cE} \right)=0\,.
\label{conmg}
\ee
The quantity $\bar{\cU}$ depends on $t$ through 
the metrics $\gamma^{ij}=a^{-2}(t)\delta^{ij}$ and 
$f_{ij}=b^2(t)\delta_{ij}$. Computing the time derivative 
$\dot{\bar{\cU}}=(\partial \bar{\cal U}/\partial \gamma^{ij})
\dot{\gamma}^{ij}+(\partial \bar{\cal U}/\partial f_{ij})
\dot{f}_{ij}$, it follows that 
\be
\dot{\bar{\cU}}+6\bar{N}H\,{\rm d}\cU
=6\bar{N}H_b\,{\rm d}{\cal U}\,,
\label{con1}
\ee
where 
\be
H_b \equiv \frac{\dot{b}}{\bar{N}b}
\ee
is the expansion rate of $b(t)$ in the metric $f_{ij}$.
From Eqs.~(\ref{conmg}) and (\ref{con1}) we obtain 
\be
H \left( 2{\rm d} \cE-\bar{\cE} \right)=
-2\bar{N}H_b\,{\rm d}{\cal U}\,.
\label{HcE}
\ee

Since $P_{\rm mg}=M_{\rm pl}^2m^2[
2(1-H_b/H){\rm d}{\cal U}-\bar{\cal U}]$
in the expanding Universe ($H>0$), using Eqs.~(\ref{back1}) and (\ref{back2}) leads to the following relation 
\be
\frac{2M_{\rm pl}^2 \dot{H}}{\bar{N}}
=-2M_{\rm pl}^2 m^2 \left( 1-r_b 
\right){\rm d}\cU-\bar{\rho}_M-\bar{P}_M\,,
\label{dotH}
\ee
where 
\be
r_b \equiv \frac{H_b}{H}\,.
\label{rb}
\ee
If the massive gravity sector is responsible for the late-time 
cosmic acceleration, the equation of state in the dark sector 
is given by 
\be
w_{\rm DE} \equiv \frac{P_{\rm mg}}{\rho_{\rm mg}}
=-1+2\left( 1-r_b
\right) \frac{{\rm d}\cU}{\bar{\cU}}\,,
\label{wde}
\ee
so that the accelerated expansion occurs for 
$(1-r_b){\rm d}\cU/\bar{\cU}<1/3$.

For the theories in which $b$ is constant, we have 
$H_b=0$ and hence $r_b=0$ for $H \neq 0$. 
In this case we have $2{\rm d}{\cal E}-\bar{{\cal E}}=0$ 
from Eq.~(\ref{HcE}), so the energy density $\rho_{\rm mg}$ 
and the pressure $P_{\rm mg}$ depend 
on $\bar{\cU}$ and ${\rm d}\cU$ alone 
with the equation of state 
$w_{\rm DE}=-1+2{\rm d}\cU/\bar{\cU}$ \cite{Comelli3}.
Then the de Sitter solution ($w_{\rm DE}=-1$) 
exists only for ${\rm d}\cU=0$, but in this case 
the theory is plagued by the strong coupling problem 
in which two of kinetic terms in the second-order action 
of perturbations vanish (as we will see later 
in Sec.\,\ref{masssec}).
The dynamical dark energy scenario with 
$w_{\rm DE}$ different from $-1$ is realized for 
the time-varying function ${\rm d}{\cal U}/\bar{{\cal U}}$ 
different from $0$, in which case the problem of vanishing 
kinetic terms can be avoided.

If $b$ is not a constant, which is the case for massive gravity 
with dilaton-like symmetry, the de Sitter solution can be 
realized for $r_b=1$ \cite{Lan}. 
In this case the expansion rate $H_b$ is identical to $H$, 
so that $2{\rm d} \cE-\bar{\cE}=-2\bar{N}{\rm d}\cU$ and 
$P_{\rm mg}=-M_{\rm pl}^2m^2 \bar{\cU}
=-\rho_{\rm mg}$. Then the quantity ${\rm d}\cU$ does 
not need to vanish on this de Sitter solution, so it 
is not plagued by the strong coupling problem 
of vanishing kinetic terms.
 
We will consider a more general situation in which 
$r_b$ is not necessarily equivalent to 1.
We require the condition that the quantity ${\rm d}\cU$ 
does not vanish from the past 
(below the strong coupling scale 
$\sim \sqrt{m M_{\rm pl}}$) to today. 
As we will see in Sec.\,\ref{consec} for concrete models, 
it is possible to realize the dynamical dark energy scenario 
with $w_{\rm DE}>-1$ for $r_b<1$ and 
$w_{\rm DE}<-1$ for $r_b>1$.
The stability of such cosmological solutions against 
ghost and Laplacian instabilities will be discussed 
in Sec.\,\ref{staconsec}.

\subsection{Concrete models}
\label{consec}

We consider concrete models of the $SO(3)$-invariant massive gravity given by the graviton potential (\ref{Vdef}) 
with the two functions \cite{Lan}
\ba
\cU &=& u_0+u_1{\cal K}^{ij}f_{ij}+\frac{1}{2} u_{2a} 
\left( {\cal K}^{ij} f_{ij} \right)^2+\frac{1}{2}u_{2b} 
{\cal K}^{ij}f_{jk}{\cal K}^{kl}f_{li}\,,\label{calU}\\
\cE &=& {\cal F}(t) 
\left[ v_0+v_1\gamma^{ij}f_{ij}+\frac{1}{2} v_{2a} 
\left( \gamma^{ij} f_{ij} \right)^2+\frac{1}{2}v_{2b} 
\gamma^{ij}f_{jk}\gamma^{kl}f_{li}
+\frac{1}{2} w_2 \xi^i \xi^j f_{ij} \right]\,,\label{calE}
\ea
where $u_0, u_1, u_{2a}, u_{2b}$ and 
$v_0, v_1, v_{2a}, v_{2b}, w_2$ are constants, and 
${\cal F}(t)$ is a function of $t$.
{}From Eq.~(\ref{UE0}) the quantity $\cE$ can contain the 
time-dependence through $\phi$ in the unitary gauge, so  
the time-dependent function ${\cal F}(t)$ has been 
taken into account in Eq.~(\ref{calE}). 
On using the property 
\be
\bar{\gamma}^{ij}f_{jk}=Y^2 \delta^i_k\,,\qquad 
Y \equiv \frac{b}{a}\,,
\ee
and the definition (\ref{dUdef}), 
the background values of $\cU, \cE$ and the 
derivatives $\rd \cU, \rd \cE$ are given, 
respectively, by 
\ba
& &
\bar{\cU}=u_0+3u_1Y^2+\frac{3}{2} u_2 Y^4\,,\qquad
\bar{\cE}={\cal F}(t) 
\left( v_0+3v_1Y^2+\frac{3}{2} v_2 Y^4 
\right)\,, \nonumber \\
& &
\rd \cU=u_1Y^2+u_2Y^4\,,\qquad 
\rd \cE={\cal F}(t)  \left(v_1Y^2+v_2Y^4 \right)\,,
\label{cUde}
\ea
where 
\be
u_2 \equiv 3u_{2a}+u_{2b}\,,\qquad
v_2 \equiv 3v_{2a}+v_{2b}\,.
\label{u2v2}
\ee

We study the background cosmological dynamics in the presence of nonrelativistic matter with energy density 
$\bar{\rho}_m$ and pressure $\bar{P}_m=0$ 
as well as radiation with energy density 
$\bar{\rho}_r$ and pressure $\bar{P}_r=\bar{\rho}_r/3$. 
Then, the background Eqs.~(\ref{back1}) 
and (\ref{dotH}) yield
\ba
3M_{\rm pl}^2H^2 &=& M_{\rm pl}^2 m^2 
\left( u_0+3u_1Y^2+\frac{3}{2} u_2 Y^4 \right)
+\bar{\rho}_m+\bar{\rho}_r\,,\label{back1n}\\
\frac{2M_{\rm pl}^2 \dot{H}}{\bar{N}}
&=&
-2M_{\rm pl}^2 m^2 (1-r_b) \left( u_1Y^2+u_2Y^4 
\right)-\bar{\rho}_m-\frac{4}{3} \bar{\rho}_r\,.
\label{back2n}
\ea
{}From Eq.~(\ref{HcE}) there is the relation 
\be
v_0+v_1Y^2-\frac12 v_2 Y^4
=2\frac{\bar{N}(t)}{{\cal F}(t)}r_b 
\left( u_1 Y^2+u_2Y^4 \right)\,.
\label{backcon0}
\ee
We choose the function ${\cal F}(t)$ to have 
the same time dependence as $\bar{N}(t)$, i.e., 
\be
{\cal F}(t)=\bar{N}(t)\,.
\label{FNcon}
\ee
If the time $t$ in Eq.~(\ref{backmet}) plays the role of 
standard cosmic time, the choice (\ref{FNcon}) 
simply corresponds to ${\cal F}(t)=\bar{N}(t)=1$. 
The constraint (\ref{backcon0}) reduces to
\be
v_0+\left( v_1-2r_b u_1 \right)Y^2
-\left( \frac{1}{2} v_2+2r_b u_2 
\right) Y^4=0\,.
\label{backcon}
\ee

Introducing the density parameters
\be
\Omega_{{\rm DE}0} \equiv \frac{m^2u_0}{3H^2}\,,
\qquad 
\Omega_{{\rm DE}1} \equiv \frac{m^2u_1Y^2}{H^2}\,,
\qquad 
\Omega_{{\rm DE}2} \equiv \frac{m^2u_2Y^4}{2H^2}\,,
\qquad 
\Omega_{m} \equiv \frac{\bar{\rho}_m}{3M_{\rm pl}^2H^2}\,,
\qquad
\Omega_r \equiv \frac{\bar{\rho}_r}{3M_{\rm pl}^2H^2}\,,
\label{dendef}
\ee
we can express Eqs.~(\ref{back1n}) and (\ref{back2n}) 
in the forms 
\ba
\Omega_m
&=& 1-\Omega_{{\rm DE}0}-\Omega_{{\rm DE}1}
-\Omega_{{\rm DE}2}-\Omega_r\,,
\label{auto0}\\
\mu 
&\equiv& \frac{\dot{H}}{\bar{N} H^2}
= -\left( 1-r_b \right)  \left( \Omega_{{\rm DE}1}
+2\Omega_{{\rm DE}2} \right)
-\frac{3}{2} \Omega_m-2\Omega_r\,.
\label{mueq}
\ea
To avoid the negative dark energy density, 
we will focus on the case in which 
$\Omega_{{\rm DE}0}, \Omega_{{\rm DE}1}, \Omega_{{\rm DE}2}$ are all positive, i.e., 
\be
u_0 \geq 0\,,\qquad u_1 \geq 0\,,\qquad u_2 \geq 0\,.
\ee
The dark energy equation of state (\ref{wde}) reduces to
\be
w_{\rm DE}=-1+\frac{2}{3} \left( 1-r_b \right) 
\frac{\Omega_{{\rm DE}1}+2\Omega_{{\rm DE}2}}
{\Omega_{{\rm DE}0}+\Omega_{{\rm DE}1}
+\Omega_{{\rm DE}2}}\,.
\label{wdees}
\ee
Defining the e-folding number $x \equiv \ln a$, 
the density parameters obey the differential equations
\ba
\frac{d\Omega_{{\rm DE}0}}{dx}
&=& -2\mu \Omega_{{\rm DE}0}\,,
\label{auto1} \\
\frac{d\Omega_{{\rm DE}1}}{dx}
&=& 2 \left( r_b-1-\mu \right)\Omega_{{\rm DE}1}\,,\\
\frac{d\Omega_{{\rm DE}2}}{dx}
&=& 2 \left( 2r_b-2-\mu \right)\Omega_{{\rm DE}2}\,,
\label{auto3}\\
\frac{d\Omega_{r}}{dx}
&=& -\left( 4+2\mu \right)\Omega_r\,.
\label{auto4}
\ea

In the following, we will consider theories in which 
$r_b$ is constant. Since the two scale factors $a$ and $b$ 
are related to each other as $b \propto a^{r_b}$, it follows that 
\be
Y=Y_0a^{r_b-1}\,,
\label{Yevo}
\ee
where $Y_0$ is constant.
Since the evolution of $Y$ is different depending 
on the values of $r_b$, we will discuss the three cases 
(1) $r_b=1$, (2) $r_b<1$, and (3) $r_b>1$, separately. 
Apart from the case (1) we choose the constant $Y_0$ 
in Eq.~(\ref{Yevo}) to be 1 without loss of generality, 
so the value of $Y$ today ($a=1$) is equivalent to 1. 

\begin{figure}[h]
\begin{center}
\includegraphics[height=3.4in,width=3.7in]{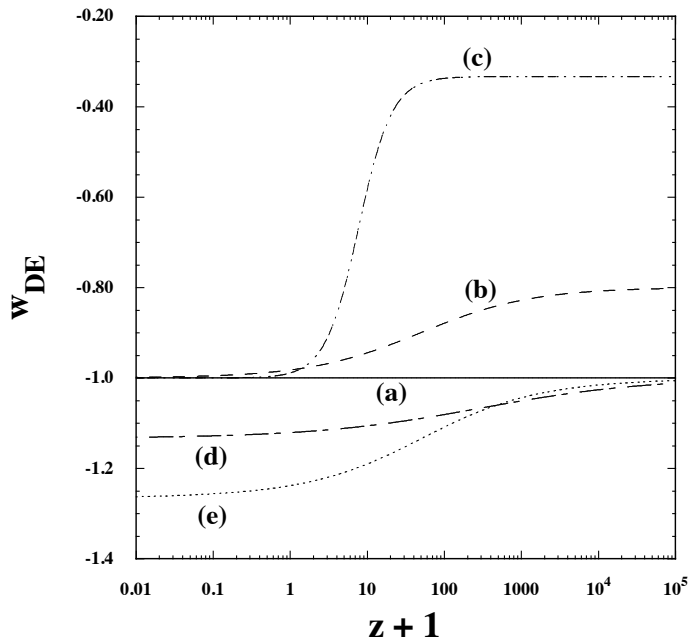}
\end{center}
\caption{\label{fig1}
Evolution of the dark energy equation of state $w_{\rm DE}$
versus $z+1=1/a$ for the models 
(a) $r_b=1$, 
(b) $r_b=0.7$, $u_0=1.86H_0^2/m^2$, 
$u_1=5.97 \times 10^{-2}H_0^2/m^2$, $u_2=0$, 
(c) $r_b=0$, $u_0=2.01H_0^2/m^2$, 
$u_1=1.11 \times 10^{-2}H_0^2/m^2$, $u_2=0$, 
(d) $r_b=1.2$, $u_0=1.89 \times 10^{-1}H_0^2/m^2$, 
$u_1=6.17 \times 10^{-1}H_0^2/m^2$, $u_2=0$, 
(e) $r_b=1.2$, $u_0=7.13 \times 10^{-2}H_0^2/m^2$, 
$u_1=9.72 \times 10^{-2}H_0^2/m^2$, $u_2=1.12H_0^2/m^2$.
The present epoch ($a=1$ and $H=H_0$) is identified 
according to the condition $\Omega_m=0.32$ with 
$\Omega_r \simeq 10^{-4}$. 
In cases (b) and (c) the dark energy equation of state is 
in the region $w_{\rm DE}>-1$, whereas in cases 
(d) and (e) the phantom equation of state is realized.}
\end{figure}

\subsubsection{$r_b=1$}

For $r_b=1$ the expansion rates $H_b$ and 
$H$ are equivalent to each other, so the ratio 
$Y=b/a$ remains to be a constant value $Y_0$. 
Then, Eq.~(\ref{backcon}) gives a constraint among the coefficients 
$v_0, v_1, v_2, u_1, u_2$, i.e., 
\be
v_0+\left( v_1 -2u_1 \right) Y_0^2
-\left( \frac12 v_2+2u_2 \right)Y_0^4=0\,.
\ee
Since $Y$ is constant, the terms $u_1 Y^2$ and 
$u_2Y^4$ in Eq.~(\ref{back1n}) behave as a cosmological 
constant with their vanishing contributions 
to the r.h.s. of Eq.~(\ref{back2n}). 
At the background level, the model with $r_b=1$ is 
equivalent to the $\Lambda$CDM model
($w_{\rm DE}=-1$), see
case (a) in Fig.~\ref{fig1}.

\subsubsection{$r_b<1$}

For $r_b<1$, the quantity $Y$ varies in proportion to 
$a^{r_{b}-1}$. Since the relation (\ref{backcon}) needs to hold for arbitrary values of $a$, the coefficients $v_0, v_1, v_2$ are subject to constraints
\be
v_0=0\,,\qquad v_1=2r_b u_1\,,\qquad 
v_2=-4r_b u_2\,.
\label{v0con}
\ee
The quantities $u_1Y^2$ and $u_2Y^4$ in 
Eqs.~(\ref{back1n}) and (\ref{back2n}) decrease with the 
growth of $a$, so the system approaches the de Sitter 
solution driven by the constant $u_0$. 
The fixed point of  the dynamical system 
(\ref{auto1})-(\ref{auto4}) corresponding to this asymptotic 
de Sitter solution is given by 
\be
{\text{(A)}}~~
(\Omega_{{\rm DE}0}, \Omega_{{\rm DE}1}, 
\Omega_{{\rm DE}2},\Omega_r, \Omega_m)=(1,0,0,0,0)\,,
\quad \text{with} \quad \mu=0\,,\quad w_{\rm DE}=-1\,.
\ee
The density parameters $\Omega_{{\rm DE}1}$ 
and $\Omega_{{\rm DE}2}$ dominate over 
$\Omega_{{\rm DE}0}$ in the asymptotic past ($a \to 0$). 
In this regime, $w_{\rm DE}$ is given by  
\be
w_{\rm DE} (a \to 0)=
  \begin{cases}
    -\left( 1+2r_b \right)/3&  \text{for $\Omega_{\rm DE1} 
    \gg \Omega_{\rm DE2}$}\,,\\
     +\left( 1-4r_b \right)/3&  \text{for $\Omega_{\rm DE1} 
    \ll \Omega_{\rm DE2}$}\,,
  \end{cases}
  \label{wdeas}
\ee
both of which are larger than $-1$ for $r_b<1$. 
Since the quantity $Y$ evolves in time, the existence of 
terms $\Omega_{{\rm DE}1}$ and $\Omega_{{\rm DE}2}$ 
in Eq.~(\ref{wdees}) leads to a dynamical dark energy 
scenario in which $w_{\rm DE}$ starts to evolve from 
the value (\ref{wdeas}) and then it finally approaches the 
asymptotic value $-1$. 
In this case, the dark energy equation of state is always 
in the region $w_{\rm DE}>-1$. 
In case (b) of Fig.~\ref{fig1} we show the evolution of 
$w_{\rm DE}$ for $r_b=0.7$ and $u_2=0$ (i.e., $\Omega_{\rm DE2}=0$), 
in which case the initial value of $w_{\rm DE}$ is 
$-(1+2r_b)/3=-0.8$ and the solution finally approaches the 
de Sitter fixed point (A).

For the theories with $r_b=0$ (i.e., $b={\rm constant}$)  
the asymptotic values (\ref{wdeas}) reduce to 
$w_{\rm DE}=-1/3$ for $\Omega_{\rm DE1} \gg 
\Omega_{\rm DE2}$ and 
$w_{\rm DE}=1/3$ for $\Omega_{\rm DE1} \ll
\Omega_{\rm DE2}$. 
Note that, for $r_b=0$, the coefficients $v_0, v_1, v_2$ are 
constrained to be 0 from Eq.~(\ref{v0con}). 
In case (c) of Fig.~\ref{fig1} we plot the evolution of  
$w_{\rm DE}$ for $r_b=0$ and $u_2=0$, which shows 
that $w_{\rm DE}$ evolves from the value close to 
$-1/3$ and asymptotically approaches $-1$. 

The quantity ${\rm d}{\cal U}m^2=(u_1Y^2+u_2Y^4)m^2$, 
which appears in the second-order 
perturbed action discussed later in Sec.~\ref{masssec}, 
is associated with the strong coupling scale 
$\Lambda_{\rm SC} \sim \sqrt{mM_{\rm pl}({\rm d}{\cal U})^{1/2}}$, 
see Sec.\,\ref{vecsec}.
For $r_b<1$, $\Lambda_{\rm SC}$ decreases with 
the growth of $a$.
For $u_1$ and $u_2$ not much less than unity
the today's value of ${\rm d}{\cal U}$ is not significantly smaller than 1, 
so the strong coupling scale today is as high as $\sqrt{mM_{\rm pl}}$.
The strong coupling problem arises only in the asymptotic future at 
which ${\rm d}{\cal U}$ sufficiently approaches 0.
This property is different from the self-accelerating solution in dRGT 
theory in which two coefficients of kinetic terms 
exactly vanish \cite{Emir}. 

\subsubsection{$r_b>1$}

For $r_b>1$ the quantities $u_1Y^2$ and $u_2Y^4$ increase 
with the growth of $a$, so the Universe finally enters the regime in which these terms dominate over the constant $u_0$ 
in Eq.~(\ref{back1n}). Note that there is the relation (\ref{v0con}) 
among the coefficients $v_0,v_1,v_2,u_1,u_2$.
Let us first discuss the theories satisfying
\be
u_2=0\,,
\ee
in which case $v_2=0$. 
Then, the solutions finally approach the fixed point 
\be
{\text{(B)}}~~(\Omega_{{\rm DE}0}, \Omega_{{\rm DE}1}, 
\Omega_{{\rm DE}2},\Omega_r, \Omega_m)=(0,1,0,0,0)\,,
\quad \text{with} \quad \mu=r_b-1\,,\quad w_{\rm DE}=
-\frac{1}{3}(1+2r_b)\,.
\ee
Since $\mu>0$ and $w_{\rm DE}<-1$, 
the Hubble parameter grows according to the relation 
$H \propto Y \propto a^{r_b-1}$. 
Then, the expanding solution associated with the fixed 
point (B) is given by 
\be
a \propto \left( t_s-t \right)^{\frac{1}{1-r_b}}\,,\qquad 
H=\frac{1}{(r_b-1)(t_s-t)}\,,
\ee
where $t_s$ is a constant. 
The Hubble parameter exhibits the divergence at $t=t_s$, 
which correspond to a big-rip singularity. 
If the condition $\Omega_{{\rm DE}0} \gg \Omega_{{\rm DE}1}$ 
is satisfied in the early Universe, $w_{\rm DE}$ starts to evolve from 
the value close to $-1$ and then it finally approaches the 
phantom equation of state $-(1+2r_b)/3$. 
The case (d) plotted in Fig.~\ref{fig1} corresponds 
to $r_b=1.2$, in which case the asymptotic value of 
$w_{\rm DE}$ is $-1.13$. 
If $\Omega_{{\rm DE}0} \ll \Omega_{{\rm DE}1}$ initially, 
$w_{\rm DE}$ is always close to $-(1+2r_b)/3$.

If we consider the theories with 
\be
u_2 \neq 0\,,
\ee
the term $u_2Y^4$ in Eq.~(\ref{back1n}) finally dominates over 
the other terms. The associated fixed point is given by 
\be
{\text{(C)}}~~(\Omega_{{\rm DE}0}, \Omega_{{\rm DE}1}, 
\Omega_{{\rm DE}2},\Omega_r, \Omega_m)=(0,0,1,0,0)\,,
\quad \text{with} \quad \mu=2r_b-2\,,\quad w_{\rm DE}=
\frac{1}{3}(1-4r_b)\,,
\label{fixc}
\ee
so that $\mu>0$ and $w_{\rm DE}<-1$.
We note that $w_{\rm DE}$ at the point (C) is smaller than 
that at the point (B).
The Hubble parameter obeys $H \propto Y^2 \propto a^{2(r_b-1)}$ 
on the fixed point (C), so the integrated solutions read
\be
a \propto \left( t_s-t \right)^{\frac{1}{2(1-r_b)}}\,,\qquad 
H=\frac{1}{2(r_b-1)(t_s-t)}\,,
\ee
which exhibit the big-rip singularity at $t=t_s$. 
The early evolution of $w_{\rm DE}$ is different depending on 
which density parameters dominate over the others in 
Eq.~(\ref{wdees}). If $\Omega_{{\rm DE}0}$ is initially 
much larger than $\Omega_{{\rm DE}1}$ and 
$\Omega_{{\rm DE}2}$, then $w_{\rm DE}$ starts to 
evolve from the value close to $-1$ and finally approaches 
$(1-4r_b)/3$. If there is a period in which $\Omega_{{\rm DE}1}$
gets larger than the other density parameters, $w_{\rm DE}$ temporally 
approaches the value $-(1+2r_b)/3$. 
In case (e) plotted in Fig.~\ref{fig1}, which corresponds to $r_b=1.2$,  
$w_{\rm DE}$ starts to evolve from the value close to $-1$ 
and approaches the asymptotic value $-1.27$.

For $r_b>1$ the quantity ${\rm d}{\cal U}m^2=(u_1Y^2+u_2Y^4)m^2$ 
increases with the growth of $a$, 
so it goes to 0 in the asymptotic past. 
As we will see in Sec.\,\ref{staconsec}, 
the decrease of ${\rm d}{\cal U}m^2$ toward the past is not 
significant for $r_b$ close to 1.
Then, the strong coupling scale $\Lambda_{\rm SC} \sim \sqrt{mM_{\rm pl}({\rm d}{\cal U})^{1/2}}$ in the past is not very different from 
its today's value.
In the asymptotic future the quantity ${\rm d}{\cal U}m^2$ goes to 
infinity with the approach to the big-rip singularity, around which 
the perturbative analysis breaks down.

\vspace{0.3cm}

The above discussion shows that the Lorentz-violating 
model given by the functions (\ref{calU}) and (\ref{calE}) 
allows for a variety of the dark equation of state: 
(1) $w_{\rm DE}=-1$ for $r_b=1$, 
(2) $w_{\rm DE}>-1$ for $r_b<1$, and 
(3) $w_{\rm DE}<-1$ for $r_b>1$, during the past 
cosmic expansion history.
In Sec.\,\ref{staconsec} we will study whether 
this model is free from the problems of ghosts 
and Laplacian instabilities.

\section{Second-order linearized action of 
the graviton potential}
\label{masssec}

In this section, we expand the action (\ref{Smg}) with the graviton 
potential given by Eq.~(\ref{Vdef}) up to second order 
in perturbations on the flat FLRW background (\ref{backmet}). 
The ADM line element (\ref{ADM}) can 
accommodate the perturbation $\delta N$ in the lapse 
function, as $N=\bar{N}(t)+\delta N$, with the 
shift perturbation $N^i$. 
We introduce the perturbation $\delta g_{ij}$ in the 
three-dimensional metric $\gamma_{ij}$, as 
\be
\gamma_{ij}=a^2(t) \left( \delta_{ij}+\delta g_{ij} 
\right)\,.\label{gammaij}
\ee
Since the unitary gauge is chosen, the perturbation 
$\delta \phi$ in the scalar field $\phi$ vanishes. 
This means that the dependence of $f_{ij}=b^2(\phi) \delta_{ij}$ in 
$\cU$ and $\cE$ does not generate any perturbed quantity 
after varying the action $S_{\rm mg}$. 
The action (\ref{Smg}) can be expressed as 
\be
S_{\rm mg}=-M_{\rm pl}^2 m^2 \int d^4 x 
\sqrt{\gamma} \left( N \cU+\cE-\cU_i \cU_{ij}^{-1} 
\cE_j \right)\,,
\label{Smgcon}
\ee
where $\gamma$ is the determinant of the three-dimensional ADM metric $\gamma_{ij}$. In the following, we expand the 
action (\ref{Smgcon}) up to second order in perturbations 
for the functions $\cU=\cU({\cal K}^{ij})$ and $\cE=\cE(\gamma^{ij}, \xi^i)$.  
In doing so, we use the following properties 
\ba
\cU_i=-2\rd \cU \bar{\gamma}_{ik}\xi^k+{\cal O}(\epsilon^2)\,,\qquad 
\cU_{ij}=-2\rd \cU \bar{\gamma}_{ij}+{\cal O}(\epsilon)\,,\qquad 
\cE_j={\rm d}^2 \cE\,\bar{\gamma}_{ij} \xi^i
+{\cal O}(\epsilon^2)\,, 
\label{Uire}
\ea
where $\epsilon$ describes the order of perturbations, and 
\be
\overline{\frac{\partial^2 \cE}{\partial \xi^i \partial \xi^j}}
\equiv {\rm d}^2 \cE\,\bar{\gamma}_{ij}\,.
\label{Edef2}
\ee
The last test term in Eq.~(\ref{Smgcon}) is of second order, 
i.e., $\cU_i \cU_{ij}^{-1} \cE_j=\rd^2 \cE
\bar{\gamma}_{ij} \xi^i \xi^j$. 
On using Eqs.~(\ref{NiN}) and (\ref{Uire}), the vector $\xi^i$ 
is related to $N^i$, as 
\be
\xi^i=\frac{2\rd \cU}{2\bar{N}\rd \cU+\rd^2 \cE}N^i
+{\cal O}(\epsilon^2)\,.
\ee
If $\rd^2 \cE=0$, then $\xi^i=N^i/\bN+{\cal O}(\epsilon^2)$. 

The square root of the determinant $\gamma$ contains the contributions of zero-th order, first order, and second order, as 
\be
\left( \sqrt{\gamma} \right)^{(0)}=a^3\,,\qquad 
\left( \sqrt{\gamma} \right)^{(1)}=\frac{1}{2}a^3 
\delta g_i^i\,,\qquad 
\left( \sqrt{\gamma} \right)^{(2)}=\frac{1}{8}a^3 
\left( \delta g_i^i \delta g_j^j
-2 \delta g_{ij} \delta g^{ij} \right)\,,
\ee
where $\delta g^i_i \equiv \delta^{ij}\delta g_{ij}$.
In the action $S_{\rm mg}$, there is the 
second-order contribution 
\be
S_{\cal U} \equiv -M_{\rm pl}^2m^2 \int d^4x 
\left( N\sqrt{\gamma} \right)^{(2)}\,\bar{\cU}
=-\int d^4x  \left( N\sqrt{\gamma} \right)^{(2)}\,
\rho_{\rm mg}\,, 
\ee
which we will separate from other contributions. 
Up to second order in perturbations, the quantities $\cU$ 
and $\cE$ are expanded as 
\ba
\cU &=& \bar{\cU}+{\rm d}\cU\,\bar{\gamma}_{ij} 
\delta \gamma^{ij} -{\rm d}\cU \bar{\gamma}_{ij} 
\xi^i \xi^j+\frac{1}{2}\overline{\frac{\partial^2 \cU}
{\partial {\cal K}^{ij}\partial {\cal K}^{kl}}}
\delta \gamma^{ij}\delta \gamma^{kl}+{\cal O}(\epsilon^3)\,,\\
\cE &=& \bar{\cE}+{\rm d}\cE\,\bar{\gamma}_{ij} 
\delta \gamma^{ij} +\frac{1}{2}{\rm d}^2\cE \bar{\gamma}_{ij} 
\xi^i \xi^j+\frac{1}{2}\overline{\frac{\partial^2 \cE}
{\partial \gamma^{ij}\partial \gamma^{kl}}}
\delta \gamma^{ij}\delta \gamma^{kl}+{\cal O}(\epsilon^3)\,.
\ea
The second-order term containing the contribution 
$\delta N^2$ arises only from the action $S_{\cal U}$. {}From the action $-M_{\rm pl}^2m^2 \int d^4x
(N\sqrt{\gamma})^{(1)}{\cal U}$ in $S_{\rm mg}$,
there is the second-order
product  $-M_{\rm pl}^2 m^2 \int d^4 x\,a^3 {\rm d}\cU 
\bar{\gamma}_{ij} \delta N \delta \gamma^{ij}$ containing the perturbation $\delta N$. 
We also employ the fact that the perturbation of the 
three-dimensional ADM metric $\gamma^{ij}$ 
is written in the form 
\be
\delta \gamma^{ij}=a^{-2}(t) \left(-\delta g^{ij}
+\delta g^{ik}{\delta g_{k}}^{j} \right)+{\cal O} (\epsilon^3)\,.
\ee
On using the notations 
\ba
\overline{\frac{\partial^2 \cU}
{\partial {\cal K}^{ij}\partial {\cal K}^{kl}}} &\equiv&
\rd^2 \cU_s \bar{\gamma}_{ij} \bar{\gamma}_{kl}
+\frac{1}{2} \rd^2 \cU_t \left( \bar{\gamma}_{ik} 
\bar{\gamma}_{jl}+\bar{\gamma}_{il} \bar{\gamma}_{jk} \right)\,,\label{Us}\\
\overline{\frac{\partial^2 \cE}
{\partial \gamma^{ij}\partial \gamma^{kl}}} &\equiv&
\rd^2 \cE_s \bar{\gamma}_{ij} \bar{\gamma}_{kl}
+\frac{1}{2} \rd^2 \cE_t \left( \bar{\gamma}_{ik} 
\bar{\gamma}_{jl}+\bar{\gamma}_{il} \bar{\gamma}_{jk} \right)\,,\label{Es}
\ea
the resulting second-order action of $S_{\rm mg}$ is given by 
\ba
S_{\rm mg}^{(2)}&=&M_{\rm pl}^2m^2 \int d^4x \,\bar{N}a^3 
\biggl[ \frac{\rd \cU}{{\bar N}} \delta N \delta g^i_i
+\frac{2\rd \cU^2}{2\bar{N}\rd \cU+\rd^2\cE}\frac{a^2}{\bar{N}} 
\delta_{ij}N^i N^j+\frac12 \left( -\frac{\bar{\cE}}{4\bar{N}}+\rd \cU+\frac{\rd \cE}{{\bar N}}
-\rd^2 \cU_s-\frac{\rd^2 \cE_s}{\bar{N}} \right)
\delta g^i_i \delta g^j_j \nonumber \\
& &+\left( \frac{\bar{\cE}}{4\bar{N}}-\rd \cU-\frac{\rd \cE}{{\bar N}}
-\frac{1}{2} \rd^2 \cU_t-\frac{\rd^2 \cE_t}{2\bar{N}} \right) 
\delta g_{ij}\delta g^{ij} \biggr]
-\int d^4x (N\sqrt{\gamma})^{(2)}\,\rho_{\rm mg}\,.
\ea

If we choose 
\be
\bar{N}=a\,,
\ee
the time $t$ plays the role of conformal time $\eta$ 
with the background line element 
\be
\overline{ds}^2=a^2(\eta) 
\left(-d\eta^2+\delta_{ij}dx^i dx^j \right)\,.
\label{linecon}
\ee
We also write the four-dimensional perturbed 
metric $g_{\mu \nu}$ in the form 
\be
g_{\mu \nu}=a^2(\eta) \left( \eta_{\mu \nu}
+\delta g_{\mu \nu} \right)\,,
\label{gper}
\ee
so that the perturbations $\delta g_{00}$ and $\delta g_{0i}$ 
are related to $\delta N$ and $N^i$ according to the relations 
$\delta g_{00}=-2\delta N/\bar{N}$ and 
$\delta g_{0i}=\delta_{ij}N^j$.
Then, the second-order action in terms of the conformal 
time can be expressed as \cite{Comelli3,Lan}
\be
S_{\rm mg}^{(2)}=
M_{\rm pl}^2 \int d\eta d^3x\,
\frac{a^4}{4} \left(m_0^2\delta g_{00}^2+2m_1^2\delta g_{0i}\delta g_{0i}-2m_4^2\delta g_{00}\delta g_{ii}
+m_3^2\delta g_{ii}\delta g_{jj}
-m_2^2\delta g_{ij}\delta g_{ij} \right) 
-\int d^4x (N\sqrt{\gamma})^{(2)}\,\rho_{\rm mg}\,,
\label{Smgf}
\ee
where the terms with same subscripts are summed over, 
and\footnote{The quantities $c_i$ used in the action (56) 
of Ref.~\cite{Lan} are related to our mass terms, as 
$m_0^2=c_0m^2$, 
$m_1^2=2c_2m^2$, $m_4^2=c_1m^2$, 
$m_3^2=4c_3m^2$, and 
$m_2^2=-4c_4m^2$.} 
\ba
& &
m_0^2=0\,,\qquad 
m_1^2=\frac{4\bN \rd \cU^2}
{2\bN \rd \cU+\rd^2 \cE}m^2\,,\qquad
m_4^2=\rd \cU\,m^2\,, \nonumber \\
& & 
m_3^2=2\left( -\frac{\bar{\cE}}{4\bN}
+\rd \cU+\frac{\rd \cE}{\bN}
-\rd^2 \cU_s-\frac{\rd^2 \cE_s}{\bN} \right)m^2\,,
\qquad
m_2^2=-4\left( \frac{\bar{\cE}}{4\bN}-\rd \cU
-\frac{\rd \cE}{\bN}-\frac{1}{2} \rd^2 \cU_t
-\frac{\rd^2 \cE_t}{2\bN} 
\right)m^2,
\label{massterm}
\ea
with $\bN=a$.
The fact that $m_0^2=0$ is attributed to the absence of the 
sixth ghost DOF, which is guaranteed by the construction 
of the graviton potential of the form 
(\ref{Vdef}) \cite{Comelli1,Comelli2}. 
The four mass terms $m_1^2, m_4^2, m_3^3, m_2^3$, 
besides $m^2$ appearing in $\rho_{\rm mg}$, 
affect the dynamics of linear perturbations. 
In Secs.\,\ref{tenvecsec} and \ref{scasec} 
we will derive the equations of 
motion of tensor (two DOFs), vector (two DOFs), and scalar 
(1 DOF) perturbations in the presence of a matter perfect fluid 
by using the second-order action (\ref{massterm}).

The Lorentz-invariant FP theory \cite{FP} 
is given by the Lagrangian 
$L_{\rm FP}=(M_{\rm pl}^2m^2/4)
(-h_{\mu \nu}h^{\mu \nu}+h_{\mu}^{\mu}h_{\nu}^{\nu})$, 
where $h_{\mu \nu}$ is the perturbed part of the 
four-dimensional metric $g_{\mu \nu}$. 
The linear expansion on the Minkowski background leads to the particular relations 
$m_0^2=0$ and $m_1^2=m_4^2=m_3^2=m_2^2
=m^2$ \cite{Rubakov}, 
which is also the case for the dRGT massive 
gravity \cite{dRGT}. 
For the self-accelerating branch of dRGT theory  
the three masses $m_0^2, m_1^2, m_4^2$ vanish with 
the relation $m_3^2=m_2^2$ \cite{Emir}, so there is 
a strong coupling problem. Indeed, the ghost instability 
arises at nonlinear level on the anisotropic 
cosmological background \cite{DGM}.

In the following, we will focus on the Lorentz-violating 
massive gravity theories in which $m_1^2, m_4^2, m_3^2, 
m_2^3$ are not generally equivalent to each other. 
As shown in Ref.~\cite{Rubakov}, the difference between 
these mass terms allows for the absence of the vDVZ discontinuity. In some works of Lorentz-violating massive gravity \cite{Dub3,Beb}, the dependence of 
$X=g^{\mu \nu} \partial_{\mu}\phi \partial_{\nu}\phi$ 
was taken into account in the graviton potential 
for realizing the kinetically driven cosmic acceleration.
Since this generally gives rise to a ghost state associated with nonvanishing $m_0^2$, we will not include such 
dependence in $V$.

\section{Tensor and vector perturbations}
\label{tenvecsec}

To study the propagation of five DOFs present 
in massive gravity theories given by the action (\ref{Lag}), 
we consider the perturbations $\delta g_{\mu\nu}$ 
in the form (\ref{gper}) on the flat FLRW background.
The linear perturbation equations of motion follow 
by expanding the action (\ref{Lag}) up to quadratic 
perturbed order. 
The corresponding second-order action can be 
expressed in the form 
\be
S^{(2)}=S_{\rm EH}^{(2)}+S_{\rm mg}^{(2)}
+S_M^{(2)}\,,
\label{S2total}
\ee
where $S_{\rm EH}^{(2)}$ and $S_M^{(2)}$ 
are the second-order contributions to $S_{\rm EH}$ and 
$S_M$, respectively. 
The quadratic action arising from the graviton 
potential is given by Eq.~(\ref{Smgf}).

We decompose the perturbations in terms of irreducible representations of the $SO(3)$ group and study the perturbations in tensor, vector, and scalar sectors separately. 
The general perturbed line element on the flat FLRW background (\ref{linecon}) is given by \cite{Bardeen}
\be
ds^2=a^2(\eta) \left\{ -(1+2A) d\eta^2+2 
\left(B_{|i}+S_i \right) d\eta dx^i 
+\left[(1+2\psi) \delta_{ij}+2E_{|ij}
+2F_{i|j}+h_{ij} \right] dx^idx^j \right\}\,,
\label{permet}
\ee
where the subscript ``${}_|$'' is the covariant derivative  
with respect to the three-dimensional spatial 
metric $g_{ij}$. 
The scalar perturbations are characterized by the four 
quantities $A, B, \psi, E$. The vector perturbations are 
given by $S_i$ and $F_i$, both of which satisfy 
the transverse conditions 
\be
{S_i}^{|i}=0\,,\qquad {F_i}^{|i}=0\,.
\label{vectra}
\ee
The tensor perturbation $h_{ij}$ obeys the transverse 
and traceless conditions 
\be
{h_{ij}}^{|j}=0\,,\qquad {h_i}^i=0\,.
\label{tentra}
\ee

For the computation of the second-order matter 
action $S_M^{(2)}$, we decompose the 
temporal and spatial components 
of $J^{\mu}$ in Eq.~(\ref{SM}) of the form 
\ba
J^{0} =  \mathcal{N}_{0}+\delta J\,,\qquad
J^{i} =\frac{1}{a^2(\eta)}\,\delta^{ik}
\left( \partial_{k}\delta j+W_k \right)\,,
\label{elldef}
\ea
where $\delta J$ and $\delta j$ correspond to 
scalar perturbations, and 
$\partial_{k} \delta j \equiv \partial \delta j/\partial x^k$.
The vector perturbation $W_k$ obeys the transverse 
condition $\partial^k W_k=0$. This is satisfied 
for the choice
\be
W_k=(W_1(\eta,z),W_2(\eta,z),0)\,,
\label{Wk}
\ee
where $z$ is the third component of the spatial 
vector $x^i$. 
We decompose the scalar quantity $\ell$ in the form 
\be
\ell=-\int^{\eta} a\bar{\rho}_{M,n} 
(\tilde{\eta})d\tilde{\eta} 
-\bar{\rho}_{M,n}v\,,
\label{ells}
\ee
where $\bar{\rho}_{M,n} \equiv 
\partial \bar{\rho}_M/\partial n$ and 
$v$ is the velocity potential. At the background level 
we have $\partial \bar{\ell}/\partial \eta=-a\bar{\rho}_{M,n}$, 
so the matter pressure (\ref{PMva}) reduces to 
\be
\bar{P}_M=n_0 \bar{\rho}_{M,n}-\bar{\rho}_M\,,
\label{PMcon}
\ee
where we used $\bar{N}=a$.

For the vector mode, we can choose
${\cal A}_1,{\cal A}_2,{\cal B}_1,{\cal B}_2$ 
in the following forms \cite{Procacosmo}
\ba
& &
{\cal A}_1=\delta {\cal A}_1(\eta,z)\,,\qquad  
{\cal A}_2=\delta {\cal A}_2(\eta,z)\,,\qquad 
{\cal B}_1=x+\delta {\cal B}_1(\eta,z)\,,\qquad
{\cal B}_2=y+\delta {\cal B}_2(\eta,z)\,,\label{ABi}
\ea
where $\delta {\cal A}_{1,2}$ and $\delta {\cal B}_{1,2}$ 
are perturbed quantities that depend on $\eta$ and $z$. 
Varying the matter action (\ref{SM}) with respect to 
$J^{\mu}$, it follows that 
\be
u_{\mu} \equiv \frac{J_{\mu}}{n\sqrt{-g}}
=\frac{1}{\bar{\rho}_{M,n}} \left( \partial_{\mu} \ell
+{\cal A}_1 \partial_{\mu}{\cal B}_1
+{\cal A}_2 \partial_{\mu}{\cal B}_2 \right)\,,
\label{umu}
\ee
where $u_{\mu}$ corresponds to the four velocity.
Substituting Eq.~(\ref{ells}) into (\ref{umu}), the spatial 
component of $u_{\mu}$ can be expressed as 
\be
u_i=-\partial_i v+v_i\,,
\ee
where the vector components $v_i$ (with $i=1,2$) are related to $\delta {\cal A}_i$, as
\be
\delta {\cal A}_i=\bar{\rho}_{M,n}v_i\,.
\label{AiVi}
\ee
The transverse condition $\partial^iv_i=0$ is satisfied for 
the choice of ${\cal A}_i$ given in Eq.~(\ref{ABi}). 

In this section, we compute the second-order actions of tensor 
and vector perturbations and derive conditions for the 
absence of ghosts and Laplacian instabilities.

\subsection{Tensor Perturbations}

Let us first compute the second-order action of tensor perturbations. The simplest choice of $h_{ij}$ obeying the 
transverse and traceless conditions (\ref{tentra}) 
is given by 
\be
h_{11}=a^2(\eta) h_1(\eta,z)\,,\qquad 
h_{22}=-a^2(\eta) h_1(\eta,z)\,,\qquad
h_{12}=h_{21}=a^2(\eta) h_2(\eta,z)\,,
\ee
where the functions $h_1$ and $h_2$ characterize 
the two propagating DOFs in the tensor sector.

The second-order tensor action arising from the 
Einstein-Hilbert term and the graviton potential
are given, respectively, by 
\ba
S_{\rm EH}^{(2)}
&=& \int d\eta d^3 x M_{\rm pl}^2 a^2 \sum_{i=1}^{2} 
\left[ \frac{1}{4} h_i'^2+\frac{1}{4}h_i \nabla^2 h_i
-\frac{1}{2} \left( 2{\cal H}'+{\cal H}^2 \right)h_i^2 
\right]\,,\\
S_{\rm mg}^{(2)}
&=& \int d\eta d^3 x a^4 \sum_{i=1}^{2}
\left(- \frac{1}{2}M_{\rm pl}^2 m_2^2 h_i^2
+\frac{1}{2} \rho_{\rm mg} h_i^2 \right)\,,
\ea
where $\nabla^2=\delta^{ij}\partial_{i}\partial_{j}$, 
${\cal H} \equiv a'/a$, and 
a prime represents the derivative with respect to $\eta$.

For the matter sector we need to expand the action 
$S_M=-\int d^4 x \sqrt{-g}\,\rho_M(n)$ up to second 
order in $h_{ij}$. The corresponding quadratic action 
can be split into the form 
$S_M^{(2)}=-\int d^4 x [(\sqrt{-g})^{(2)} \bar{\rho}_M
+\sqrt{-\bar{g}}\bar{\rho}_{M,n}\delta n]$, where 
$(\sqrt{-g})^{(2)}=-a^4(h_1^2+h_2^2)/2$ 
and $\delta n=n_0 (h_1^2+h_2^2)/2$ 
with $\sqrt{-\bar{g}}=a^4$. Then, it follows that 
\be
S_{M}^{(2)}=-\int d\eta d^3 x \sum_{i=1}^{2} 
\frac{1}{2}a^4 \bar{P}_M h_i^2\,,
\ee
where $\bar{P}_M$ is the matter pressure given by Eq.~(\ref{PMcon}).

The background equations follow from Eqs.~(\ref{back1}) 
and (\ref{back2}) by replacing $H$ with ${\cal H}/a$ 
and setting $\bar{N}=a$, i.e., 
\ba
& &
3M_{\rm pl}^2 {\cal H}^2=a^2 \left( 
\rho_{\rm mg}+\bar{\rho}_M \right)\,,
\label{back1d}\\
& &
M_{\rm pl}^2 \left( 2{\cal H}'+{\cal H}^2 \right)
=-a^2 \left( P_{\rm mg}+\bar{P}_M \right)\,.
\label{back2d}
\ea
On using Eqs.~(\ref{back1d})-(\ref{back2d}) and the relation 
\be
\rho_{\rm  mg}+P_{\rm mg}
=2M_{\rm pl}^2 m^2 \left( 1-r_b \right) 
{\rm d}\cU\,,
\ee
the total second-order action (\ref{S2total}) 
in the tensor sector reads 
\be
S_T^{(2)}=\int d\eta d^3x \frac{M_{\rm pl}^2}{2} 
\sum_{i=1}^2 
\left( \frac12 a^2 h_i'^2+\frac12a^2h_i \nabla^2 h_i
-\frac12 a^4m_T^2 h_i^2 \right)\,,
\label{ST2}
\ee
where 
\be
m_T^2 \equiv 2m_2^2-4m^2\left( 1-r_b \right) 
{\rm d}\cU\,.
\label{mT}
\ee

The first two terms in Eq.~(\ref{ST2}) are exactly the same as those in GR, so there is neither ghost nor Laplacian instability
in the tensor sector. The effect of massive gravity arises 
only through the effective graviton mass squared $m_T^2$. 
If $w_{\rm DE}=-1$, i.e., $r_b=1$ or ${\rm d}\cU=0$, 
then the effective tensor mass squared (\ref{mT}) is equivalent to $2m_2^2$. The deviation from $w_{\rm DE}=-1$ 
leads to the modification to the value $2m_2^2$.

The fact that the kinetic and gradient terms remain unchanged 
relative to those in GR is an important property of 
the $SO(3)$-invariant massive gravity. The tensor propagation speed $c_T$ is equivalent to 
the speed of light $c$. 
This is consistent with the recent tight bound on 
the propagation speed of gravitational waves 
constrained from GW170817 \cite{LIGO}. The mass 
$m_T$ relevant to the late-time cosmic acceleration 
is not much different from the today's Hubble expansion rate
$H_0 \sim 10^{-33}$ eV, so it also satisfies the upper limit 
$m_T \le 7.7 \times 10^{-23}$ eV constrained from 
GW170104 \cite{LIGO2}.

Varying the action (\ref{ST2}) with respect to $h_{i}$, 
the resulting equation of motion in Fourier space 
with the comoving wavenumber $k$ is given by 
\be
h_i''+2{\cal H}h_i'+ \left( k^2+a^2m_T^2 
\right)h_i=0\,.
\ee
If $m_T^2<0$, then the tachyonic instability is present 
for the modes $k^2/a^2<|m_T^2|$. 
Provided that $|m_T|$ is of the same order as 
$H_0$, the tachyonic instability is harmless for tensor perturbations 
inside the today's Hubble radius.  

\subsection{Vector Perturbations}
\label{vecsec}

The perturbed line element (\ref{permet}) contains the vector perturbations $S_i$ and $F_i$.
We choose them in the forms 
$S_i=(S_1(\eta, z), S_2(\eta, z),0)$ and 
$F_i=(F_1(\eta, z), F_2(\eta, z),0)$ to satisfy the transverse conditions (\ref{vectra}). 
Then, the quadratic actions $S_{\rm EH}^{(2)}$ 
and $S_{\rm mg}^{(2)}$ of the vector sector are given, respectively, by 
\ba
S_{\rm EH}^{(2)} &=&
\int d\eta d^3 x M_{\rm pl}^2 a^2 \sum_{i=1}^{2} 
\left[ -\frac{1}{4}  
(S_i-F_i')\nabla^2 (S_i-F_i')
+\frac{3}{2}{\cal H}^2S_i^2 +
\frac12 (2{\cal H}'+{\cal H}^2)F_i \nabla^2 F_i \right]\,,\\ 
S_{\rm mg}^{(2)} &=&
\int d\eta d^3 x\,a^4 \sum_{i=1}^{2} \left[
\frac{1}{2} M_{\rm pl}^2 \left( m_1^2 S_i^2
+m_2^2F_i \nabla^2 F_i \right)
-\frac{1}{2} \rho_{\rm mg} \left( 
S_i^2+F_i \nabla^2 F_i \right) \right]\,.
\ea
The vector perturbations $W_i,{\cal A}_i,{\cal B}_i$ 
in the Schutz-Sorkin action are chosen to be of the forms (\ref{Wk}) and (\ref{ABi}).
Then, the second-order matter action reads 
\be
S_M^{(2)} =\int d\eta d^3 x \sum_{i=1}^{2} 
\biggl[ \frac{1}{2a^3{\cal N}_0} \left\{ 
\bar{\rho}_{M,n} (W_i+{\cal N}_0 a^2 S_i)^2 
-{\cal N}_0a^7 \bar{\rho}_M S_i^2 \right\}
-{\cal N}_0 \delta {\cal A}_i \delta {\cal B}_i'
-\frac{1}{a^2}W_i \delta {\cal A}_i +\frac12 a^4 
\bar{P}_MF_i \nabla^2 F_i
\biggr]\,.
\label{SMs}
\ee
The perturbations $W_i, \delta {\cal A}_i,  \delta {\cal B}_i$ appear only in the action $S_M^{(2)}$. 
Varying Eq.~(\ref{SMs}) with respect to $W_i$ and 
using Eq.~(\ref{AiVi}), it follows that 
\be
W_i=a{\cal N}_0 (v_i-aS_i)\,.
\label{maeq1}
\ee
Variations of the action (\ref{SMs}) with respect to 
$\delta {\cal A}_i$ and $\delta {\cal B}_i$ lead, 
respectively, to 
\ba
v_i &=& a(S_i-\delta {\cal B}_i' )\,,\label{vire}\\
\delta {\cal A}_i &=& 
\frac{a^3(\bar{\rho}_M+\bar{P}_M)}
{{\cal N}_0}v_i=C_i\,,
\label{maeq3}
\ea
where $C_i$ are constants. 
Substituting Eqs.~(\ref{maeq1})-(\ref{maeq3}) into 
Eq.~(\ref{SMs}), the second-order matter 
action reduces to 
\be
S_M^{(2)}=\int d\eta d^3 x \sum_{i=1}^{2} 
\left[ \frac{1}{2}a^2 (\bar{\rho}_M+\bar{P}_M) 
v_i^2- \frac{1}{2}a^4 \bar{\rho}_M S_i^2+\frac{1}{2}a^4\bar{P}_M F_i \nabla^2 F_i 
\right]\,.
\ee
On using the background Eqs.~(\ref{back1d})-(\ref{back2d}), the total second-order 
action of vector perturbations reduces to 
\be
S_V^{(2)} =\int d\eta d^3 x \sum_{i=1}^{2}
\biggl[ -\frac{1}{4} M_{\rm pl}^2 a^2 
(S_i-F_i')\nabla^2 (S_i-F_i')+ \frac{1}{2}a^2 
(\bar{\rho}_M+\bar{P}_M) v_i^2 
+\frac12 M_{\rm pl}^2a^4 m_1^2 S_i^2 
+\frac14 M_{\rm pl}^2a^4m_T^2
F_i \nabla^2 F_i
\biggr]\,,
\label{S2vec}
\ee
where $m_T^2$ is defined by Eq.~(\ref{mT}).

Taking note that $v_i$ is related to $S_i$ through Eq.~(\ref{vire}) and varying the action (\ref{S2vec}) with respect to $S_i$, we obtain
\be
M_{\rm pl}^2 \nabla^2 (S_i-F_i')
-2M_{\rm pl}^2a^2 m_1^2S_i
-\frac{2{\cal N}_0}{a^2}C_i=0\,.
\ee
The Fourier components of $S_i$ corresponding to 
the comoving wavenumber $k$ obey
\be
S_i=\frac{M_{\rm pl}^2a^2k^2 F_i'-2{\cal N}_0C_i}
{M_{\rm pl}^2a^2 (k^2+2a^2 m_1^2)}\,.
\ee
Substituting this relation into Eq.~(\ref{S2vec}), the resulting 
second-order action is given by 
\be
S_V^{(2)}=\int d\eta d^3 x \sum_{i=1}^{2} \left[ 
\frac{M_{\rm pl}^2}{2}a^4q_V (F_i'^2-c_V^2k^2 F_i^2)
+\frac{{\cal N}_0^2 C_i^2\{M_{\rm pl}^2(k^2+2a^2 m_1^2)
+2a^2(\bar{\rho}_M+\bar{P}_M)\}}
{2M_{\rm pl}^2a^4(\bar{\rho}_M
+\bar{P}_M)(k^2+2a^2 m_1^2)} 
\right]\,,\label{S2Vf}
\ee
where 
\be
q_V=\frac{k^2 m_1^2}{k^2+2a^2 m_1^2}\,,
\qquad 
c_V^2=\frac{m_T^2}{2m_1^2} \left( 1
+\frac{2a^2 m_1^2}{k^2} \right)\,.
\label{qvcv}
\ee

For the theories satisfying $m_1^2 \neq 0$, there are two dynamical fields $F_1$ and $F_2$ with the propagation 
speed squared $c_V^2$. 
In the small-scale limit characterized by $k^2/a^2 \gg m_1^2$, 
the quantities (\ref{qvcv}) reduce to 
$q_V \simeq m_1^2$ and 
$c_V^2 \simeq m_T^2/(2m_1^2)$. 
As long as the two conditions 
\be
m_1^2>0\,,\qquad 
m_T^2 \geq 0
\label{veccon}
\ee
are satisfied, there are neither ghosts nor Laplacian instabilities for the modes $k^2/a^2 \gg m_1^2$. 
Under Eq.~(\ref{veccon}), the conditions $q_V>0$ 
and $c_V^2 \geq 0$ hold for any value of $k$. 
The two quantities $q_V$ and $c_V$ coincide 
with those derived in Ref.~\cite{Lan} in the absence 
of matter, so the presence of matter does not 
substantially modify the stability conditions of 
vector perturbations. 
The naive expectation is that adding the Lorentz-invariant Schutz-Sorkin action to the Lorentz-violating graviton 
action might give rise to some nontrivial modification to 
the kinetic and gradient terms of vector perturbations.
We have explicitly shown that this is not the case. 

It is worth mentioning that there is an associated strong coupling scale of the propagating vector fields $F_i$. This becomes apparent after normalizing the kinetic terms $M_{\rm pl}^2 q_VF_i'^2/2$ to 
canonical forms. Since $q_V \simeq m_1^2$
for the modes $k^2/a^2 \gg m_1^2$, the strong 
coupling scale $\Lambda_{\rm SC}$ will be of the order 
$\sqrt{m_1M_{\rm pl}}$. 
In other words, our low-energy effective theory and the 
resulting cosmological solutions are valid up to the scale $\sqrt{m_1M_{\rm pl}}$. 
If we consider the theory with ${\rm d}^2 {\cal E}=0$, 
which is realized for $w_2=0$ in the function (\ref{calE}), 
then we have $m_1^2=2{\rm d}{\cal U}m^2$ and 
$\Lambda_{\rm SC} \sim \sqrt{m M_{\rm pl}({\rm d}{\cal U})^{1/2}}$. 
For the dynamical dark energy scenario in which 
${\rm d}{\cal U}$ varies
in time, the scale $\Lambda_{\rm SC}$ is time-dependent. 
In Sec.\,\ref{staconsec} we will discuss the variation of 
$\Lambda_{\rm SC}$ 
in concrete models of the late-time cosmic acceleration.

\section{Scalar perturbations}
\label{scasec}

We proceed to the derivation of the second-order action of 
scalar perturbations and obtain the no-ghost and stability 
conditions in the presence of a matter perfect fluid. 

\subsection{Second-order action}

With the notation of Eq.~(\ref{gammaij}), the scalar metric 
perturbations $A, B,\psi, E$ in the line element (\ref{permet}) 
can be expressed as
\be
\delta g_{00}=-2A\,,\qquad  
\delta g_{0i}=\partial_i B\,,\qquad 
\delta g_{ij}=2\psi \delta_{ij}+2\partial_{i} \partial_{j}E\,.
\ee
The covariant derivative $E_{|ij}$ in Eq.~(\ref{permet}) has been replaced with the partial derivative $\partial_{i} \partial_{j}E$ 
by reflecting the fact that the term arising from 
the Christoffel symbols $\Gamma^{i}_{jk}$ are 
at most second-order in perturbations.
The quadratic action arising from the Einstein-Hilbert 
term (\ref{SEH}) is given by 
\ba
S_{\rm EH}^{(2)}
&=& \int d \eta d^3x\,M_{\rm pl}^2 a^2 \biggl[ 
2(\psi'-{\cal H}A) \nabla^2 (B-E')-3\psi'^2-2A\nabla^2 \psi
-\psi \nabla^2 \psi+6{\cal H} A\psi'-\frac{9}{2}{\cal H}^2A^2
+9{\cal H}^2A\psi \nonumber \\
&&
+\frac{3}{2}{\cal H}^2 \left\{ 2A \nabla^2 E
+(\partial_i B)^2 \right\}+\frac{1}{2} \left\{ 3\psi^2 
+2\psi \nabla^2 E-(\nabla^2 E)^2 \right\} 
(2{\cal H}'+{\cal H}^2) \biggr]\,.
\label{Ssca1}
\ea
The second-order action arising from the graviton
potential (\ref{Smg}) yields
\ba
S_{\rm mg}^{(2)}
&=&  \int d \eta d^3x\,M_{\rm pl}^2 a^4 
\biggl[ m_0^2A^2+\frac{1}{2}m_1^2 (\partial_i B)^2
-m_2^2  \left\{ 3\psi^2 
+2\psi \nabla^2 E+(\nabla^2 E)^2 \right\} 
+m_3^2 \left( 3\psi+\nabla^2 E \right)^2  \nonumber \\
& &+2m_4^2 A \left( 3\psi+\nabla^2 E \right)
-\frac{\rho_{\rm mg}}{2M_{\rm pl}^2} \left\{ 
2A \nabla^2 E-B\nabla^2 B-A^2+6A\psi
+3\psi^2 +2\psi \nabla^2 E-(\nabla^2 E)^2 \right\}
\biggr]\,.
\label{Ssca2}
\ea
We have not omitted the term $m_0^2$ to discuss 
the ghost instability for the theories with 
$m_0^2 \neq 0$ later.
For the matter sector, we define the matter density 
perturbation $\delta \rho_M$, as
\be
\delta \rho_M \equiv \frac{\bar{\rho}_{M,n}}{a^3} 
\left[ \delta J-{\cal N}_0 \left( 3\psi+
\nabla^2 E \right) \right]\,.
\ee
On using Eq.~(\ref{elldef}), the perturbation of the number 
density (\ref{num}) expanded up to quadratic order is given by 
\be
\delta n=\frac{\delta \rho_M}{\bar{\rho}_{M,n}}
-\frac{2a^7{\cal N}_0 \delta \rho_M (3\psi+\nabla^2 E)
+\{ a^4{\cal N}_0^2[(\partial_i B)^2+(3\psi-\nabla^2 E)
(\psi+\nabla^2 E)]
+2{\cal N}_0a^2\partial_i B \partial_i \delta j
+(\partial_i \delta j)^2 \} \bar{\rho}_{M,n}}
{2a^7{\cal N}_0 \bar{\rho}_{M,n}},
\ee
which is equivalent to $\delta \rho_M/\bar{\rho}_{M,n}$ 
at linear order. The second-order action following from 
the expansion of the Schutz-Sorkin action (\ref{SM}) in 
scalar perturbations reads
\ba
S_{M}^{(2)}
&=&  \int d \eta d^3x
\biggl[ \frac{a^4 \bar{\rho}_M}{2} A^2
+\frac{a^4 \bar{P}_M}{2}
\left\{ 3\psi^2+ (\partial_i B)^2\right\}
-\frac{a^4 \bar{\rho}_{M,nn}}
{2\bar{\rho}_{M,n}^2}
\delta \rho_M^2+\frac{\bar{\rho}_{M,n}}{2a^6n_0} 
(\partial_i \delta j)^2+\frac{\bar{\rho}_{M,n}}{a^2} 
\partial_i \delta j \partial_i v+\frac{\bar{\rho}_{M,n}}{a} 
\partial_i \delta j \partial_i B \nonumber \\
& & +a^3 v' \delta \rho_M-a^4A \delta \rho_M 
-3a^3 \psi \left( 3n_0^2{\cal H} \bar{\rho}_{M,nn}v
+a\bar{\rho}_M A-n_0 \bar{\rho}_{M,n}v' \right)
-\frac{3n_0a^3{\cal H} \bar{\rho}_{M,nn}}{\bar{\rho}_{M,n}}
v \delta \rho_M \nonumber \\
& &-\frac{a^4}{2} \nabla^2 E
\left\{ n_0 \bar{\rho}_{M,n} (\nabla^2 E-2\psi)
+\bar{\rho}_M (2A+2\psi-\nabla^2 E) \right\}
+a^3 \nabla^2 E \left( n_0 \bar{\rho}_{M,n}v'
-3n_0^2\bar{\rho}_{M,nn}{\cal H}v \right) \biggr]\,,
\label{Sm2}
\ea
where $\bar{\rho}_{M,nn} \equiv \partial^2 \bar{\rho}_M
/\partial n^2|_{n=n_0}$.
Varying this action with respect to $\delta j$, it follows that 
\be
\partial_i \delta j=-a^4 n_0 \left( \partial_i v+ a\,
\partial_i B \right)\,.
\label{delj}
\ee
Substituting Eq.~(\ref{delj}) into Eq.~(\ref{Sm2}), 
the second-order matter action reduces to 
\ba
\hspace{-0.15cm}
S_{M}^{(2)}
&=&  \int d \eta d^3x
\biggl[ a^3 \left( v'-3{\cal H}c_M^2 v-aA \right) \delta \rho_M
-\frac{a^4 c_M^2}{2n_0 \bar{\rho}_{M,n}} \delta \rho_M^2
-\frac{a^2}{2}n_0 \bar{\rho}_{M,n} \left\{ (\partial_i v)^2
+2a\partial_i v \partial_i B \right\}
+\frac{a^4 \bar{\rho}_M}{2} \left( A^2-6A \psi \right)
 \nonumber \\
\hspace{-0.15cm}
& & +n_0 a^3 (v'-3c_M^2 {\cal H} v)\bar{\rho}_{M,n} 
(3\psi+\nabla^2 E)-\frac{a^4}{2} \bar{\rho}_M
\left\{ 2A \nabla^2 E+(\partial_i B)^2 \right\}
+\frac{a^4}{2} \bar{P}_M  \left\{ 3\psi^2 
+2\psi \nabla^2 E-(\nabla^2 E)^2 \right\} \biggr],
\label{Sms}
\ea
where $c_M$ is the matter sound speed defined by 
\be
c_M^2=\frac{n_0 \bar{\rho}_{M,nn}}{\bar{\rho}_{M,n}}
=\frac{\bar{P}_M'}{\bar{\rho}_M'}\,.
\ee

Taking the sum of Eqs.~(\ref{Ssca1}), (\ref{Ssca2}), (\ref{Sms}) and using the background Eqs.~(\ref{back1d})-(\ref{back2d}), the total second-order action of 
scalar perturbations is given by 
\ba
S_S^{(2)}&=&\int d\eta d^3 x 
\biggl[ M_{\rm pl}^2 a^2 
\left\{
2(\psi'-{\cal H}A) \nabla^2 (B-E')-3{\cal H}^2A^2
+6{\cal H}A\psi'-2A \nabla^2 \psi
-3\psi'^2-\psi \nabla^2 \psi  \right\} \nonumber \\
& &-a^4 A \delta \rho_M+a^3 (v'-3{\cal H}c_M^2 v)
\{\delta \rho_M+n_0 \bar{\rho}_{M,n}(3\psi+\nabla^2 E) \}
-\frac{a^4 c_M^2}{2n_0 \bar{\rho}_{M,n}} \delta \rho_M^2
+\frac{a^2}{2}n_0 \bar{\rho}_{M,n} \left( v \nabla^2 v
+2a v \nabla^2 B \right) \nonumber \\
& &+M_{\rm pl}^2 a^4 \biggl\{
m_0^2A^2-\frac{1}{2}m_1^2 B \nabla^2 B
-3m_{2+}^2\psi^2 -2m_{2+}^2\psi \nabla^2 E
-m_{2-}^2(\nabla^2 E)^2 
+m_3^2 \left( 3\psi+\nabla^2 E \right)^2 \nonumber \\
& &
+2m_4^2A \left( 3\psi+\nabla^2 E \right) \biggr\}\biggr]\,,
\label{SSfull}
\ea
where 
\ba
& &
m_{2+}^2 \equiv m_2^2+m^2 
\left(1 -r_b \right) {\rm d}{\cal U}
=m_2^2+\frac{m^2}{2} \bar{\cal U} 
\left(1+w_{\rm DE} \right)\,,\\
& &
m_{2-}^2 \equiv m_2^2-m^2
\left(1 -r_b \right) {\rm d}{\cal U}
=m_2^2-\frac{m^2}{2} \bar{\cal U} 
\left(1+w_{\rm DE} \right)\,.
\ea
If $w_{\rm DE}$ is equivalent to $-1$, then 
$m_{2+}^2=m_{2-}^2=m_2^2$. 
The deviation from $w_{\rm DE}=-1$ leads to the modification
to the mass term $m_2^2$. 
Note that there is the relation 
$3m_{2-}^2-m_{2+}^2=m_T^2$.

\subsection{Stability conditions of scalar perturbations}

Varying the action (\ref{SSfull}) with respect to $v$ and 
$\delta \rho_M$ and using the relations 
$n_0 \bar{\rho}_{M,n}=\bar{\rho}_M+\bar{P}_M$ and 
$\bar{P}_M'=c_M^2 \bar{\rho}_M'$, it follows that 
\ba
& &
\delta \rho_M'+3{\cal H} \left(1+c_M^2 \right) \delta \rho_M
+(\bar{\rho}_M+\bar{P}_M) \left( 3\psi'+\nabla^2 \sigma
-\frac{1}{a} \nabla^2 v \right)=0\,,
\label{sper1}\\
& &
v'-3{\cal H}c_M^2v-aA-\frac{ac_M^2}
{\bar{\rho}_M+\bar{P}_M} \delta \rho_M=0\,,
\label{sper2}
\ea
where 
\be
\sigma \equiv E'-B\,.
\ee
Variations of the action (\ref{SSfull}) with respect to 
$A, B, \psi, E$ lead to
\ba
\hspace{-1.2cm}
& &
3{\cal H} \left( \psi'-{\cal H} A \right)-\nabla^2 \psi
+{\cal H} \nabla^2\sigma+a^2 \left[ m_0^2 A
+m_4^2 (3\psi+\nabla^2 E) 
\right]=\frac{a^2}{2M_{\rm pl}^2} \delta \rho_M\,,
\label{sper3}\\
\hspace{-1.2cm}
& &
\psi'-{\cal H} A-\frac{1}{2}a^2 m_1^2B
=-\frac{a}{2M_{\rm pl}^2} 
\left( \bar{\rho}_M+\bar{P}_M \right)v\,,
\label{sper4}\\
\hspace{-1.2cm}
& &
\psi''+2{\cal H}\psi'-{\cal H} A'-\left( 2{\cal H}'+{\cal H}^2 
\right)A+a^2 \left[ \{ m_4^2-m^2(1-r_b){\rm d}{\cal U} \} A
+(3m_3^2-m_{2+}^2)\psi+(m_3^2-m_{2-}^2)\nabla^2 E 
\right] \nonumber \\
& &
=-\frac{a^2c_M^2}{2M_{\rm pl}^2} \delta \rho_M,
\label{sper5}\\
\hspace{-1.2cm}
& &
\sigma'+2{\cal H}\sigma-A-\psi+a^2m_T^2 E=0\,,
\label{sper6}
\ea
where we used Eq.~(\ref{sper2}).

In the following, we switch to the Fourier space with the comoving wavenumber $k$. To discuss the evolution of matter density perturbations, we introduce the gauge-invariant density contrast
\be
\delta \equiv \frac{\delta \rho_M}{\bar{\rho}_M}
+3\left( 1+\frac{\bar{P}_M}{\bar{\rho}_M} \right) \psi\,.
\ee
We express Eq.~(\ref{sper1}) in terms of $\delta$ and $\delta'$
and then solve it for $v$. 
The Hamiltonian and momentum constraints (\ref{sper3}) and 
(\ref{sper4}) are also used to eliminate the perturbations 
$A$ and $B$ from the action (\ref{SSfull}).
The kinetic term appearing in the second-order scalar 
action can be written in the form 
$S_K^{(2)}=\int d\eta d^3 x \vec{\cal X}^{t'}{\bm K}
\vec{\cal X}'$, where $\vec{\cal X}^{t}
=\left( \delta/k, kE, \psi \right)$ and ${\bm K}$ is 
a $3 \times 3$ matrix. The three matrix components 
$K_{11}, K_{22}, K_{33}$ of ${\bm K}$ 
are given, respectively, by 
\ba
K_{11}&=& 
\frac{a^4M_{\rm pl}^2[a^2m_1^2(3\cH^2-a^2m_0^2)+2\cH^2k^2]}
{2a^2 (\bar{\rho}_M+\bar{P}_M+M_{\rm pl}^2m_1^2)
(3\cH^2-a^2m_0^2)+4M_{\rm pl}^2 \cH^2k^2} 
\frac{\bar{\rho}_M^2}{\bar{\rho}_M+\bar{P}_M}\,, \\
K_{22} &=&\frac{a^4 M_{\rm pl}^2 
[a^2( \bar{\rho}_M+\bar{P}_M)(3 \cH^2-a^2 m_0^2)
+2M_{\rm pl}^2 \cH^2 k^2]}
{2a^2 (\bar{\rho}_M+\bar{P}_M+M_{\rm pl}^2m_1^2)
(3\cH^2-a^2 m_0^2)+4M_{\rm pl}^2 \cH^2 k^2}m_1^2\,,\\
K_{33}&=&\frac{a^4 M_{\rm pl}^2 
[3a^2( \bar{\rho}_M+\bar{P}_M+M_{\rm pl}^2m_1^2)
+2M_{\rm pl}^2 k^2]}
{a^2(3 \cH^2-a^2m_0^2)(\bar{\rho}_M+\bar{P}_M
+M_{\rm pl}^2m_1^2)+2M_{\rm pl}^2 \cH^2k^2}m_0^2 \,.
\label{K33}
\ea
Provided that $m_1^2 \neq 0$ and $m_0^2 \neq 0$, there are 
three dynamical propagating fields $\delta, E, \psi$. 
To see the ghost instability arising from the nonvanishing $m_0^2$, 
let us consider the situation without matter. 
In this case, the two fields $E$ and $\psi$ propagate and 
the nonvanishing matrix components are 
\be
K_{22}=\frac{M_{\rm pl}^2m_1^2a^4\cH^2k^2}
{a^2m_1^2(3\cH^2-a^2m_0^2)+2\cH^2k^2}\,,\qquad
K_{23}=K_{32}=-\frac{a^2m_0^2}{k \cH^2}K_{22}\,,\qquad
K_{33}=\frac{M_{\rm pl}^2m_0^2a^4(3a^2m_1^2+2k^2)}
{a^2m_1^2(3\cH^2-a^2m_0^2)+2\cH^2k^2}\,.
\ee
The absence of ghosts requires that the matrix ${\bm K}$ is 
positive definite. In the Minkowski limit (${\cal H} \to 0$), 
the no-ghost conditions translate to 
$M_{\rm pl}^4k^2a^4 <0$ and 
$M_{\rm pl}^2(3m_1^2a^2+2k^2)/m_1^2<0$. 
Since the first condition is not satisfied, the theories 
with $m_0^2 \neq 0$ are plagued by the ghost problem 
on the Minkowski background.

The $SO(3)$-invariant massive gravity given by 
the action (\ref{Lag}) satisfies the condition 
$m_0^2=0$ and hence $K_{33}=0$ in Eq.~(\ref{K33}), 
under which the field $\psi$ does not corresponds to 
a dynamical field. In this case, the field $\psi$ can be eliminated from the second-order scalar action. 
In the presence of matter, we are left with two dynamical 
fields $\delta$ and $E$ with a quadratic action in the form 
\be
S_{S}^{(2)}=\int d\eta d^3 x 
\left( \vec{\cal X}^{t'}{\bm K}
\vec{\cal X}'
+k^2 \vec{\mathcal{X}}^{t}{\bm L}
\vec{\mathcal{X}}
-\vec{\mathcal{X}}^{t}{\bm M}
\vec{\mathcal{X}}
-\vec{\mathcal{X}}^{t}{\bm B}
\vec{\mathcal{X}}'
\right) \,,
\label{L2mat}
\ee
where ${\bm K}$, ${\bm L}$, ${\bm M}$, ${\bm B}$
are $2 \times 2$ matrices, and 
\be
\vec{\mathcal{X}}^{t}=\left( \delta/k, kE \right) \,.
\ee
The matrices ${\bm M}$ and ${\bm B}$ contain the contributions up to the order of $k^0$.
In the small-scale limit ($k \to \infty$), the components 
of ${\bm K}$ are given, respectively, by 
\ba
K_{11} &=& \frac{a^4\bar{\rho}_M^2}{2(\bar{\rho}_M
+\bar{P}_M)}
+{\cal O} \left( \frac{1}{k^2} \right)\,,\\
K_{12} &=& K_{21}=\frac{3a^6 \bar{\rho}_M
[2M_{\rm pl}^2m_1^2({\cal H}'-{\cal H}^2)
+a^2\{ m_1^2 ( \bar{\rho}_M+\bar{P}_M)+4M_{\rm pl}^2 
m_4^2 (m_1^2-m_4^2)\}]}
{4[2M_{\rm pl}^2 ({\cal H}^2-{\cal H}')
-a^2(M_{\rm pl}^2m_1^2+\bar{\rho}_M
+\bar{P}_M)]}\frac{1}{k^2}
+{\cal O} \left(  \frac{1}{k^4} \right)\,,\\
K_{22} &=& \frac{a^4M_{\rm pl}^2[2M_{\rm pl}^2 
\{ ({\cal H}^2-{\cal H}')m_1^2+2a^2m_4^2(m_4^2-m_1^2)\}
-a^2m_1^2 ( \bar{\rho}_M+\bar{P}_M)]}
{4M_{\rm pl}^2 ({\cal H}^2-{\cal H}')
-2a^2(M_{\rm pl}^2m_1^2+\bar{\rho}_M+\bar{P}_M)}
+{\cal O} \left(  \frac{1}{k^2} \right)\,.
\label{K22f}
\ea
If the matrix ${\bm K}$ is positive definite, the ghost is absent. 
In the small-scale limit, this amounts to the conditions that 
the leading-order terms of both $K_{11}$ and $K_{22}$ 
are positive. The first condition corresponds to 
$\bar{\rho}_M+\bar{P}_M>0$, whereas the second 
translates to 
\be
q_S \equiv \frac{2m_4^2 (m_1^2-m_4^2)-m_1^2m^2(1-r_b)
{\rm d}{\cal U}}
{m_1^2-2m^2(1-r_b){\rm d}{\cal U}}>0\,,
\label{ghostsca}
\ee
where we used the background 
Eqs.~(\ref{back1d})-(\ref{back2d}). 
If $w_{\rm DE}=-1$, i.e., 
$(1-r_b){\rm d}{\cal U}=0$, the condition 
(\ref{ghostsca}) translates to 
$m_4^2(m_1^2-m_4^2)/m_1^2>0$.
Since we require that $m_1^2>0$ to avoid the 
vector ghost, the scalar ghost is absent under the 
condition $0<m_4^2<m_1^2$.
This condition agrees with that derived in Ref.~\cite{Lan} 
on the de Sitter background. 
{}From Eq.~(\ref{ghostsca}) we observe that the deviation from $w_{\rm DE}=-1$ affects the no-ghost condition of scalar perturbations.

The components of the matrix ${\bm L}$ in the $k \to \infty$ 
limit read
\ba
L_{11} &=& -c_M^2 \frac{a^4 \bar{\rho}_M^2}
{2(\bar{\rho}_M+\bar{P}_M)}
+{\cal O} \left(  \frac{1}{k^2} \right)\,,\\
L_{12} &=& L_{21}=
{\cal O} \left(  \frac{1}{k^2} \right)\,,\\
L_{22} &=& a^4M_{\rm pl}^2\left( m_3^2-m_{2-}^2 \right)
+{\cal O} \left(  \frac{1}{k^2} \right)\,.
\ea
The leading-order contribution to the product $\delta E$ is 
at most of the order $k^0$ in the action (\ref{L2mat}), 
so it appears only through the matrix components $M_{12}$ and $M_{21}$ in ${\bm M}$.
The scalar propagation speed squared $c_S^2$ can derived  
from the dispersion relation 
\be
{\rm det} \left( c_S^2 {\bm K}+{\bm L} \right)=0\,.
\ee
In the small-scale limit, we obtain the two solutions
\ba
c_{S1}^2 &=& c_M^2\,,\\
c_{S2}^2 &=& \frac{2(m_{2-}^2-m_3^2)[2M_{\rm pl}^2 ({\cal H}^2-{\cal H}')
-a^2(M_{\rm pl}^2m_1^2+\bar{\rho}_M+\bar{P}_M)]}
{2M_{\rm pl}^2 
\{ ({\cal H}^2-{\cal H}')m_1^2+2a^2m_4^2(m_4^2-m_1^2)\}
-a^2m_1^2 (\bar{\rho}_M+\bar{P}_M)}\,.
\label{cS2}
\ea
The Laplacian instability can be avoided for $c_M^2 \geq 0$ 
and $c_{S2}^2 \geq 0$. 
On using the background equations of motion,
 the second condition translates to 
\be
c_{S2}^2=\frac{(m_{2-}^2-m_3^2)
[m_1^2-2m^2(1-r_b){\rm d}{\cal U}]}
{2m_4^2(m_1^2-m_4^2)-m^2m_1^2 (1-r_b)
{\rm d}{\cal U}} \geq 0\,.
\label{cscon}
\ee
For $w_{\rm DE}=-1$ this condition reduces to 
$c_{S2}^2=(m_{2}^2-m_3^2)m_1^2/
[2m_4^2 (m_1^2-m_4^2)] \geq 0$, which agrees
with that derived in Ref.~\cite{Lan}.

\subsection{Stability of concrete models}
\label{staconsec}

We have shown that, in the small-scale limit,  
there are neither ghosts nor Laplacian instabilities 
for tensor, vector, and scalar perturbations under the conditions 
\ba
& &
q_V \simeq m_1^2>0\,,\label{sum1}\\
& & 
c_V^2 \simeq \frac{m_T^2}{2m_1^2}
=\frac{m_2^2-2m^2(1-r_b){\rm d}{\cal U}} {m_1^2} \geq 0\,,\\
& &
q_S=\frac{2m_4^2 (m_1^2-m_4^2)-m_1^2m^2(1-r_b)
{\rm d}{\cal U}}
{m_1^2-2m^2(1-r_b){\rm d}{\cal U}}>0\,,\\
& &
c_{S2}^2=\frac{[m_2^2-m_3^2-m^2(1-r_b)
{\rm d}{\cal U}]
[m_1^2-2m^2(1-r_b){\rm d}{\cal U}]}
{2m_4^2(m_1^2-m_4^2)-m^2m_1^2 (1-r_b)
{\rm d}{\cal U}} \geq 0\,.\label{sum4}
\ea
For $w_{\rm DE}=-1$, the above conditions coincide 
with those derived in Ref.~\cite{Lan}.
We have obtained the above stability conditions 
in the presence of matter without restricting to 
de Sitter solutions. The matter contribution 
$\bar{\rho}_M+\bar{P}_M$ appearing in 
Eqs.~(\ref{K22f}) and (\ref{cS2}) has been eliminated 
by using the background Eq.~(\ref{dotH}). 
The resulting no-ghost and stability conditions (\ref{sum1})-(\ref{sum4}) 
are expressed in terms of $m_1^2,m_2^2,m_3^2,m_4^2$ and 
$m^2(1-r_b){\rm d}{\cal U}$.

For concreteness, we consider the model with the functions 
${\cal U}, {\cal E}$ given by Eqs.~(\ref{calU})-(\ref{calE}). 
The quantities defined in Eqs.~(\ref{Edef2}), (\ref{Us}), 
and (\ref{Es}) reduce to 
\be
\rd^2 \cE={\cal F}(t)w_{2}Y^2\,,\qquad
\rd^2 \cU_s=u_{2a}Y^4\,,\qquad 
\rd^2 \cU_t=u_{2b}Y^4\,,\qquad 
\rd^2 \cE_s={\cal F}(t)v_{2a}Y^4\,,\qquad 
\rd^2 \cE_t={\cal F}(t)v_{2b}Y^4\,,\qquad 
\ee
with ${\rm d}{\cal U}=u_1Y^2+u_2Y^4$ 
and ${\rm d}{\cal E}={\cal F}(t)(v_1Y^2+v_2Y^4)$. 
By choosing the function ${\cal F}(t)$ in the form 
(\ref{FNcon}), the mass terms in Eq.~(\ref{massterm}) yield
\ba
& &
m_0^2=0\,,\qquad m_1^2=\frac{4Y^2 
(u_1+u_2Y^2)^2}{2(u_1+u_2Y^2)+w_2}m^2\,,
\qquad 
m_4^2=Y^2 (u_1+u_2Y^2)m^2\,,\nonumber \\
& &
m_3^2=\frac{1}{4} \left[ 
(8u_2-8u_{2a}+5v_2-8v_{2a})Y^4
+2(4u_1+v_1)Y^2-2v_0 \right]m^2\,,\nonumber \\
& &
m_2^2=\frac{1}{2}
\left[ (8u_2+4u_{2b}+5v_2+4v_{2b})Y^4
+2(4 u_1+v_1)Y^2-2v_0 \right]m^2\,.
\label{massterm2}
\ea

The stability of de Sitter solutions satisfying $r_b=1$ was studied in Ref.~\cite{Lan}, which showed the existence 
of the viable parameter space. 
Here, we study the stability conditions for 
\be
r_b \neq 1\,.
\ee
In this case there exist three constraints (\ref{v0con}) among 
the coefficients $v_0,v_1,v_2,u_1,u_2$. 
In what follows, we will also focus on the case $w_2=0$. 
Recalling the relation (\ref{u2v2}), there are five free parameters 
$u_0, u_1, u_2, u_{2a}, v_{2a}$ in the model under consideration.
The conditions (\ref{sum1})-(\ref{sum4}) translate to 
\ba
q_V&=&
2Y^2 \left( u_1+u_2 Y^2 \right)m^2>0\,,\label{noghost1}\\
c_V^2 &=&
\frac{(1+2r_b)u_1+[2(1-4r_b)u_2
-3(u_{2a}+v_{2a})]Y^2}{u_1+u_2Y^2} \geq 0\,,
\label{cVcon}\\
q_S&=&
Y^2 \left( u_1+u_2 Y^2 \right)m^2>0\,,
\label{noghost2}\\
c_S^2 &=&
\frac{(1+2r_b)u_1+[3(1-4r_b)u_2
-4(u_{2a}+v_{2a})]Y^2}{u_1+u_2Y^2} 
\geq 0\,,
\label{cScon}
\ea
which depend on the two parameters $u_1, u_2$ and 
the combination $u_{2a}+v_{2a}$.
Provided that 
\be
u_1>0\,,\quad \text{and} \quad u_2>0\,,
\ee
the no-ghost conditions (\ref{noghost1}) and (\ref{noghost2}) 
are satisfied. We also note that the density parameters 
$\Omega_{{\rm DE}1}$ and $\Omega_{{\rm DE}2}$ defined 
in Eq.~(\ref{dendef}) are positive under these conditions.

\begin{figure}[h]
\begin{center}
\includegraphics[height=3.4in,width=3.5in]{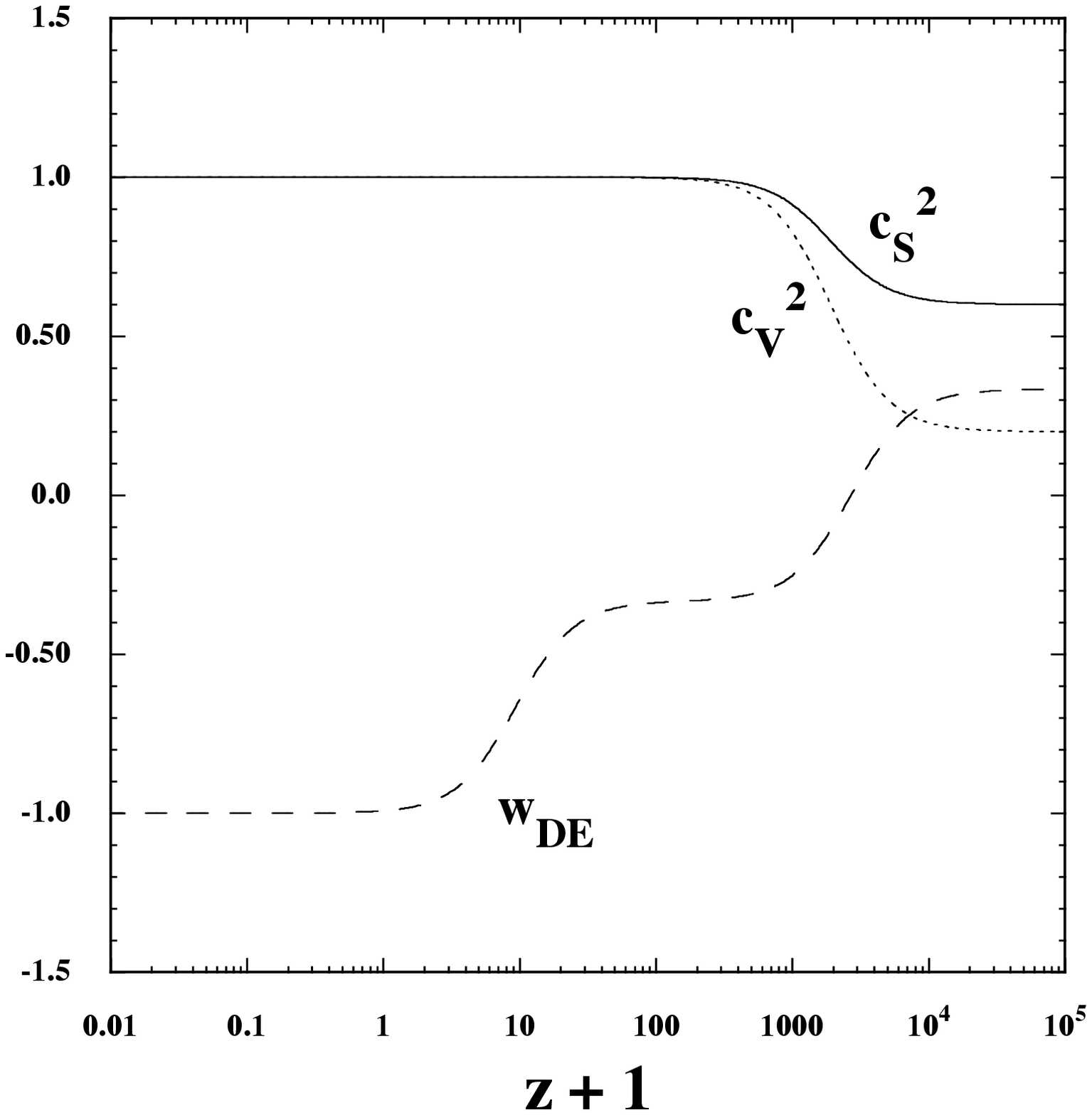}
\includegraphics[height=3.4in,width=3.5in]{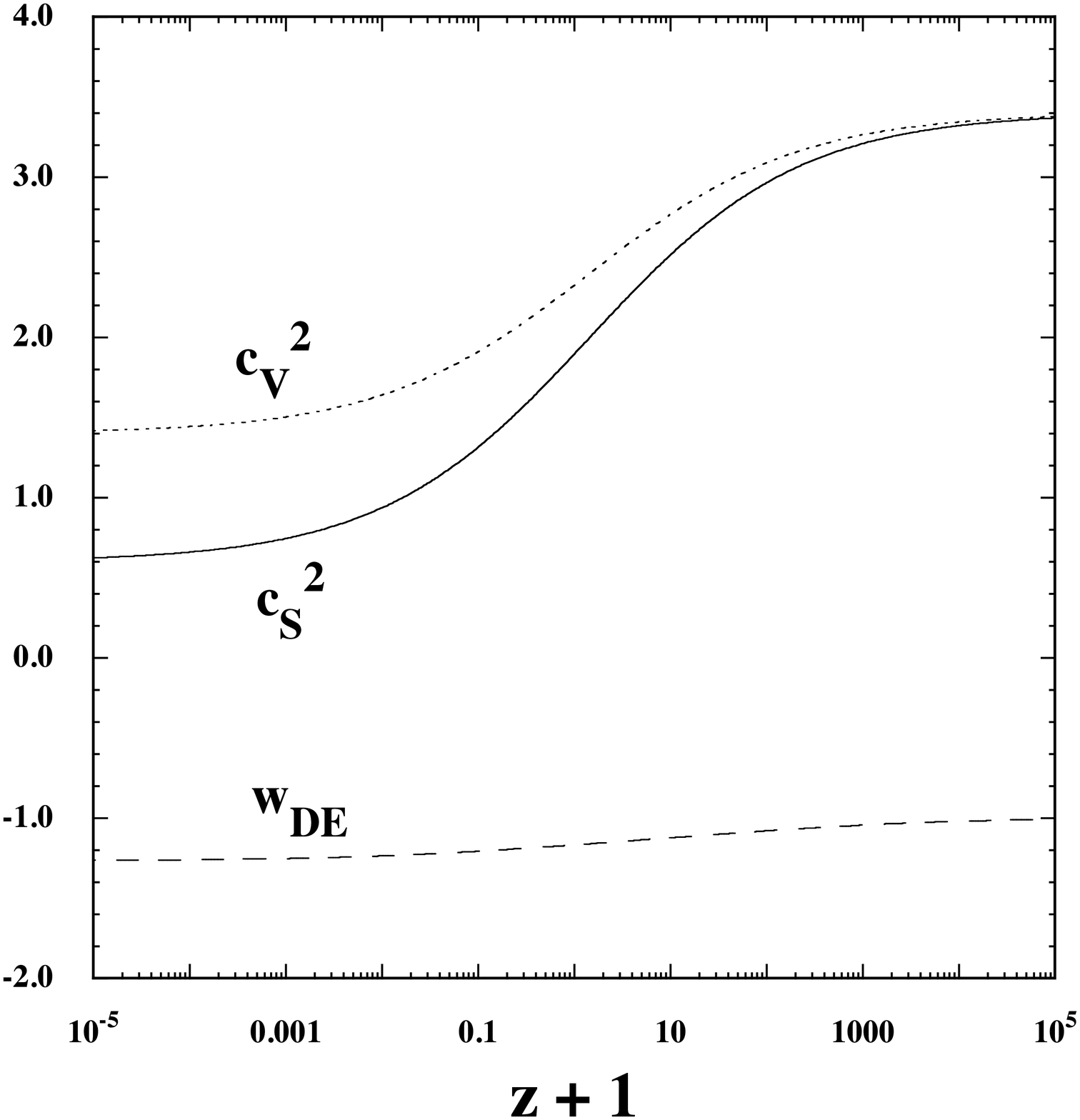}
\end{center}
\caption{\label{fig2}
Evolution of $c_V^2, c_S^2$ and $w_{\rm DE}$ versus 
$z+1=1/a$ for the models: 
(i) $r_b=0$, $u_0=2.02H_0^2/m^2$, 
$u_1=7.66 \times 10^{-3}H_0^2/m^2$, 
$u_2=2.09 \times 10^{-9}H_0^2/m^2$, 
$u_{2a}+v_{2a}=0.6u_2$ (left), 
and 
(ii) $r_b=1.2$, $u_0=1.63 \times 10^{-1} H_0^2/m^2$, 
$u_1=3.95 \times 10^{-1}H_0^2/m^2$, 
$u_2=4.61 \times 10^{-1}H_0^2/m^2$, 
$u_{2a}+v_{2a}=-3u_2$ (right).
We identify the present epoch ($a=1$ and $Y=1$) 
by the condition $\Omega_m=0.32$ 
with $\Omega_r \simeq 10^{-4}$. }
\end{figure}

Let us first discuss the case $r_b<1$, under which 
the quantity $Y=a^{r_b-1}$ decreases with the 
growth of $a$. 
In the asymptotic past ($a \to 0$ and $Y \to \infty$),
the conditions (\ref{cVcon}) and (\ref{cScon}) reduce, 
respectively, to 
\ba
c_V^2
&=& 2 \left( 1-4r_b \right)-\frac{3(u_{2a}+v_{2a})}{u_2} 
\geq 0\,,\label{cVas} \\
c_S^2
&=& 3 \left( 1-4r_b \right)-\frac{4(u_{2a}+v_{2a})}{u_2}
\geq 0 \,.\label{cSas}
\ea
If $r_b<1/4$ these conditions are satisfied for 
$(u_{2a}+v_{2a})/u_2 \le (2/3)(1-4r_b)$, whereas, 
if $r_b>1/4$, they are satisfied for 
$(u_{2a}+v_{2a})/u_2 \le (3/4)(1-4r_b)$.
In the asymptotic future ($a \to \infty$ and $Y \to 0$), 
the conditions (\ref{cVcon}) and (\ref{cScon}) translate to 
\be
c_V^2=c_S^2=1+2r_b \geq 0\,,
\label{cVSf}
\ee
which are satisfied for $r_b \geq -1/2$. 

In the left panel of Fig.~\ref{fig2} we plot the evolution of 
$c_V^2, c_S^2$ and $w_{\rm DE}$ for $r_b=0$ and 
$u_{2a}+v_{2a}=0.6u_2$. 
In this case we have $c_V^2=0.2$ and $c_S^2=0.6$ from 
Eqs.~(\ref{cVas}) and (\ref{cSas}), whereas $c_V^2=c_S^2=1$ from Eq.~(\ref{cVSf}). 
As we see in Fig.~\ref{fig2}, 
both $c_V^2$ and $c_S^2$ continuously increase from 
their past asymptotic values to the final same value 1, so 
the ghosts and Laplacian instabilities associated with 
vector and scalar perturbations are absent.
In this numerical simulation the condition 
$\Omega_{{\rm DE}1} \ll \Omega_{{\rm DE}2}$ is 
initially satisfied, so $w_{\rm DE}$ starts to evolve 
from the value close to $(1-4r_b)/3=1/3$ 
[see Eq.~(\ref{wdeas})]. 
After $\Omega_{{\rm DE}2}$ gets smaller than 
$\Omega_{{\rm DE}1}$, $w_{\rm DE}$ 
temporally stays around the value $-(1+2r_b)/3=-1/3$. 
After the dominance of $\Omega_{{\rm DE}0}$ over 
$\Omega_{{\rm DE}1}$ and $\Omega_{{\rm DE}2}$, 
$w_{\rm DE}$ approaches the asymptotic value $-1$.

For $r_b<1$, both $q_V$ and $q_S$ asymptotically approach 
$+0$ with the decrease of $Y=a^{r_b-1}$.  
Then, the strong coupling scales 
$(M_{\rm pl}\sqrt{q_{V}})^{1/2}$ and 
$(M_{\rm pl}\sqrt{q_{S}})^{1/2}$ also decrease in time. 
Provided that the today's values $\sqrt{q_V}=\sqrt{2q_S}
=\sqrt{2(u_1+u_2)}m$ are not much different from 
the order $H_0 \sim 10^{-42}$\,GeV,  the strong coupling 
scales now can be estimated as 
$\Lambda_{\rm SC} \sim \sqrt{M_{\rm pl}H_0} \sim 10^{-12}$~GeV. 
This corresponds to the energy scale 
$E_{\rm SC} \sim \sqrt{\Lambda_{\rm SC} M_{\rm pl}} \sim 10^3$~GeV, 
so our perturbative analysis is valid from 
the redshift $z \sim 10^{15}$ to today. 
In the future the strong coupling scale gradually decreases
from the today's value $\Lambda_{\rm SC} \sim 10^{-12}$~GeV, 
so the region for the validity of our perturbative analysis 
shifts to lower energy scales.

For $r_b>1$ the quantity $Y$ increases with the growth 
of $a$, so the past asymptotic values of $c_V^2$ and $c_S^2$ are given by Eq.~(\ref{cVSf}), i.e., $c_V^2=c_S^2=1+2r_b>3$.
In the asymptotic future, $c_V^2$ and $c_S^2$ approach 
the values (\ref{cVas}) and (\ref{cSas}), respectively. 
Provided that 
\be
\frac{u_{2a}+v_{2a}}{u_2} \le 
\frac{3}{4} \left(1-4r_b \right)\,,
\ee
the stability conditions of vector and scalar perturbations 
are satisfied. In the right panel of Fig.~\ref{fig2} we show 
the evolution of $c_V^2, c_S^2$ and $w_{\rm DE}$ 
for $r_b=1.2$ and $u_{2a}+v_{2a}=-3u_2$. 
In this case, $c_V^2$ and $c_S^2$ start to evolve from 
the same value $3.4$ and then they
approach the asymptotic values $c_V^2=1.4$ and 
$c_S^2=0.6$, respectively [as estimated from 
Eqs.~(\ref{cVas}) and (\ref{cSas})]. 
Initially, $\Omega_{{\rm DE}0}$ dominates over 
$\Omega_{{\rm DE}1}$ and $\Omega_{{\rm DE}2}$, 
so $w_{\rm DE}$ is close to $-1$.
Finally, the solution approaches the fixed point (C) 
given by Eq.~(\ref{fixc}) with $w_{\rm DE}=(1-4r_b)/3=-1.267$.

For $r_b>1$ the quantities $q_V$ and $q_S$ decrease 
toward the asymptotic past.
Since $q_V$ and $q_S$ are of the order $u_1 m^2Y^2$ in 
the regime $u_2Y^2 \ll u_1$, the strong coupling scales 
can be estimated as $\Lambda_{\rm SC} \sim 
\sqrt{(u_1)^{1/2}m M_{\rm pl}Y} \sim \sqrt{(u_1)^{1/2} m M_{\rm pl}}\,
a^{(r_b-1)/2}$. Provided that $r_b$ is close to 1, 
the term $a^{(r_b-1)/2}$ does not give rise to 
a significant change to $\Lambda_{\rm SC}$ 
relative to the value $\sqrt{(u_1)^{1/2}mM_{\rm pl}}$.
For $r_b=1.2$ we have $a^{(r_b-1)/2}=10^{-1}$ 
at $a=10^{-10}$, so the strong coupling scales at
the redshift $z \sim 10^{10}$ are as high as 
$\Lambda_{\rm SC} \sim 10^{-13}$~GeV for
$(u_1)^{1/2}m \sim H_0$.
The quantities $q_V$ and $q_S$, which increase toward 
the asymptotic future, exhibit divergences at the big-rip singularity. The perturbative analysis breaks down 
around the big-rip singularity.

\section{Growth of non-relativistic matter perturbations}
\label{growthsec}

We study the evolution of nonrelativistic matter perturbations and derive the effective gravitational coupling 
as well as the gravitational slip parameter associated with 
the observations of large-scale structures and weak lensing.
Let us consider non-relativistic matter satisfying 
\be
\bar{P}_M=0\,,\qquad c_M^2=0\,.
\ee
Introducing the Bardeen's gauge-invariant gravitational potentials \cite{Bardeen}
\be
\Psi \equiv A-\sigma'-{\cal H}\sigma\,,\qquad 
\Phi \equiv \psi-{\cal H}\sigma\,, 
\ee
we can express Eq.~(\ref{sper6}) in the form 
\be
\Psi+\Phi=a^2 m_T^2 E\,.
\label{anire}
\ee
In Fourier space, Eqs.~(\ref{sper1}) and (\ref{sper2}) 
reduce, respectively, to 
\ba
& &
\delta'-k^2 \sigma+k^2 \frac{v}{a}=0\,,
\label{mper1}\\
& &
v'=aA\,,
\label{mper2}
\ea
where $\delta=\delta \rho_M/\bar{\rho}_M+3\psi$.
Taking the time derivative of Eq.~(\ref{mper1}) and using 
Eq.~(\ref{mper2}), we obtain
\be
\delta''+{\cal H}\delta'+k^2 \Psi=0\,.
\label{deltaeq0}
\ee
We define the effective gravitational coupling $G_{\rm eff}$ and the gravitational slip parameter $\eta_s$, as
\ba
k^2 \Psi &=& 
-4\pi G_{\rm eff}a^2 \bar{\rho}_M \delta\,\label{Psieq},\\
\eta_s &=& -\frac{\Phi}{\Psi}\,.
\label{eta}
\ea
Provided that $m_T^2 \neq 0$, $\eta_s$ is different from 1.
The effective gravitational potential associated with the deviation of 
light rays is defined by \cite{Uzan,Sapone}
\be
\Phi_{\rm eff} \equiv \Phi-\Psi
=-\left( \eta_s+1 \right)\Psi\,.
\label{Phieff}
\ee
{}From Eqs.~(\ref{Psieq}) and (\ref{Phieff}) it follows that 
\be
k^2 \Phi_{\rm eff}=8\pi G a^2\Sigma \bar{\rho}_M \delta\,, 
\ee
where $G=1/(8\pi M_{\rm pl}^2)$ is the 
Newton's gravitational constant, and 
\be
\Sigma \equiv \frac{G_{\rm eff}}{G} 
\frac{\eta_s+1}{2}\,.
\label{Sigmadef}
\ee

Since we are interested in the growth of perturbations 
relevant to the observations of large-scale structures and weak lensing, we focus on the modes deep inside the Hubble 
radius ($k^2 \gg \cH^2$). 
For the theories satisfying the condition $m_0^2=0$ 
there are two dynamical scalar quantities $\delta$ and 
$E$, whose equations of motion follow from 
the second-order action (\ref{L2mat}). 
Varying the action (\ref{L2mat}) with respect to $\delta$ 
and $E$, the resulting equations of motion for these 
perturbations can be written in the forms
\ba
& &
\delta''+\alpha_1 \delta'+\alpha_2 \delta+\alpha_3 E'
+\alpha_4 E=0\,,
\label{deltaeq}\\
& &
E''+\beta_1 E'+\beta_2 E+\beta_3 \delta'
+\beta_4 \delta=0\,,
\label{Eeq}
\ea
where $\alpha_{1,2,3,4}$ and $\beta_{1,2,3,4}$ 
are time-dependent coefficients.
The coefficients $\alpha_{1,2,3,4}$, expanded 
around large $k$, are given by 
\ba
& &
\alpha_1={\cal H}+\frac{3\cH a^2 \bar{\rho}_M}
{2M_{\rm pl}^2}\frac{1}{k^2}
+{\cal O} \left(  \frac{1}{k^4} \right)\,,\qquad
\alpha_2=-\frac{a^2 \bar{\rho}_M}{2M_{\rm pl}^2}
+\left( \frac{3a^4 \bar{\rho}_M^2}{4M_{\rm pl}^4}
+\frac{3a^4m_4^2 \bar{\rho}_M}{2M_{\rm pl}^2} 
\right) \frac{1}{k^2}
+{\cal O}\left(  \frac{1}{k^4} \right)\,,\nonumber \\
& &
\alpha_3=-\frac{3a^2{\cal H}
[a^2 m_1^2 \bar{\rho}_M-2M_{\rm pl}^2
\{m_1^2 ({\cal H}^2-{\cal H}')+2a^2m_4^2 
(m_4^2-m_1^2) \}]}
{4M_{\rm pl}^2({\cal H}^2-{\cal H}')
-2a^2(M_{\rm pl}^2m_1^2+\bar{\rho}_M)}
+{\cal O} \left( \frac{1}{k^2} \right)\,,\nonumber \\
& & 
\alpha_4=-a^2 \left( m_{2+}^2-3m_{2-}^2+m_{4}^2 \right)k^2
+{\cal O} \left( k^0 \right)\,.
\label{albe}
\ea
For the modes $k \gg {\cal H}$, the second term
in $\alpha_1$ is much smaller than $\cH$. 
The term $3a^4\bar{\rho}_M^2/(4M_{\rm pl}^4k^2)$ in $\alpha_2$
is also negligible relative to the first contribution to $\alpha_2$.
Provided that all the mass terms $m_i^2$ as well as the combination 
$m_{2+}^2-3m_{2-}^2+m_{4}^2$ are of the similar orders to $m^2$, 
the term $\alpha_3E'$ is much smaller than $\alpha_4E$ for 
$k \gg \cH$. Then, Eq.~(\ref{deltaeq}) approximately obeys
\be
\delta''+{\cal H}\delta'-4\pi G a^2 \bar{\rho}_M \delta
\simeq a^2\left( m_{2+}^2-3m_{2-}^2+m_{4}^2 \right)
k^2E-\frac{3a^4m_4^2 \bar{\rho}_M}{2M_{\rm pl}^2} \frac{\delta}{k^2} \,,
\label{deleqmo}
\ee
where we have not ignored the term 
$3a^4m_4^2 \bar{\rho}_M/(2M_{\rm pl}^2k^2)$ 
in $\alpha_2$. 
The growth of matter perturbations is modified from that in GR
by the graviton mass terms appearing 
on the r.h.s. of Eq.~(\ref{deleqmo}).

For large $k$ the coefficients $\beta_{1,2,3,4}$ in Eq.~(\ref{Eeq}) are given by 
\ba
\hspace{-0.5cm}
& &
\beta_1={\cal O} \left( k^{0} \right)\,\qquad 
\beta_2 =c_{S2}^2\,k^2+{\cal O} \left( k^0 \right)\,,
\nonumber \\
\hspace{-0.5cm}
& &
\beta_3= {\cal O}\left( \frac{1}{k^4} \right)\,,\qquad
\beta_4=-\frac{(m_{2+}^2-3m_3^2+m_4^2)[
2M_{\rm pl}^2 ({\cal H}^2-{\cal H}')
-a^2 ( M_{\rm pl}^2m_1^2+\bar{\rho}_M)]}
{\{ [2({\cal H}^2-{\cal H'})m_1^2
+4a^2m_4^2 (m_4^2-m_1^2)]M_{\rm pl}^2
-a^2m_1^2 \bar{\rho}_M\} M_{\rm pl}^2}
\frac{a^2 \bar{\rho}_M}{k^2}
+{\cal O} \left( \frac{1}{k^4} \right)\,,
\label{betai}
\ea
where $c_{S2}^2$ is the sound speed squared defined by 
Eq.~(\ref{cS2}) with $\bar{P}_M=0$. 
The leading-order contribution to $\beta_1$, which 
does not contain the $k$-dependence, is at most 
of the order ${\cal H}$. 
Since the leading-order contribution to $\beta_3$ 
has the $k^{-4}$ dependence, the term $\beta_3 \delta'$ 
is suppressed relative to $\beta_4 \delta$ in the small-scale 
limit. Provided that the variation of $E$ is small such that 
the conditions $|E'| \lesssim |\cH E|$ and 
$|E''| \lesssim |\cH^2 E|$ are satisfied, 
the first two terms on the l.h.s. of Eq.~(\ref{Eeq})  
can be neglected relative to $\beta_2 E$ for the 
modes deep inside the sound horizon,  
\be
c_{S2}^2\,k^2 \gg \cH^2\,.
\label{subho}
\ee
In this case the perturbation $E$ is related to $\delta$ 
according to $\beta_2 E+\beta_4 \delta \simeq 0$, 
such that 
\be
E\simeq -\frac{\beta_4}{\beta_2}\delta
\simeq \frac{(m_{2+}^2-3m_3^2+m_4^2)a^2 \bar{\rho}_M}
{2(m_{2-}^2-m_3^2)M_{\rm pl}^2k^4}\delta\,.
\label{Esim}
\ee
This approximation, which is so-called the quasi-static 
approximation \cite{quasi1,quasi2,quasi3}, 
is valid for the modes satisfying the condition (\ref{subho}).
Substituting (\ref{Esim}) into Eq.~(\ref{deleqmo}), we can express the matter perturbation equation 
in the form (\ref{deltaeq0}) with Eq.~(\ref{Psieq}). 
The effective gravitational coupling is given by 
\be
G_{\rm eff}=G \left[ 1+\frac{m_4^4-(m_{2+}^2-3m_3^2+2m_4^2)m_T^2}
{m_{2-}^2-m_3^2} \frac{a^2}{k^2} \right]\,.
\label{Geffqu}
\ee
On using Eqs.~(\ref{anire}), (\ref{Psieq}), and 
(\ref{Geffqu}), the two gauge-invariant gravitational 
potentials are given by 
\ba
\Psi &=&
-\frac{4\pi G a^2 \bar{\rho}_M \delta}{k^2} 
\left[ 1+\frac{m_4^4-(m_{2+}^2-3m_3^2+2m_4^2)m_T^2}
{m_{2-}^2-m_3^2} \frac{a^2}{k^2}\right]\,,\\
\Phi &=&
\frac{4\pi G a^2 \bar{\rho}_M \delta}{k^2} 
\left[ 1+\frac{m_4^2 (m_4^2-m_T^2)}
{m_{2-}^2-m_3^2} \frac{a^2}{k^2} \right]\,.
\label{Phiqu}
\ea

In the large $k$ limit the terms containing the graviton 
masses in $G_{\rm eff}, \Psi, \Phi$ vanish, 
so the general relativistic behavior is recovered.
If all the mass terms $m_i^2$ are of the similar order 
to $m^2$, the massive gravity corrections in 
the square brackets of Eqs.~(\ref{Geffqu})-(\ref{Phiqu}) 
are at most of the order $m^2a^2/(c_{S2}^2 k^2)$. 
On using the condition (\ref{subho}), they
are much smaller than $m^2/H^2$, where $H=\cH/a$ 
is the Hubble expansion rate in terms of the 
cosmic time $t=\int a\,d\eta$.
For $m$ of the order of the today's Hubble expansion 
rate $H_0$, it follows that the graviton mass terms 
in the square brackets of Eqs.~(\ref{Geffqu})-(\ref{Phiqu}) 
are much smaller than 1 from the matter era to today.

Taking the small-scale limit, the gravitational slip parameter $\eta_s$ and the parameter $\Sigma$ reduce to 
\ba
\eta_s &\simeq& 1+\frac{(m_{2+}^2-3m_3^2+m_4^2)m_T^2}
{m_{2-}^2-m_3^2}\frac{a^2}{k^2}\,,
\label{etaqu} \\
\Sigma &\simeq& 1-\frac{(m_{2+}^2-3m_3^2+3m_4^2)m_T^2
-2m_4^4}{2(m_{2-}^2-m_3^2)}
\frac{a^2}{k^2}\,,
\label{Sigmaqu}
\ea
so that the deviations of $\eta_s$ and $\Sigma$ from 1  
are much smaller than 1 for the modes deep inside 
the sound horizon.

The above discussion is valid for the modes satisfying the 
condition (\ref{subho}). There is a specific situation in which   
$c_{S2}^2$ is much smaller than 1. 
This can be realized for $m_2^2=m_3^2$, which is the case 
for the self-accelerating solution in dRGT theory. 
Since $m_{2-}^2-m_3^2=-m^2 (1-r_b){\rm d}{\cal U}$ 
in this case, the small deviation from 
$w_{\rm DE}=-1$ leads to 
the non-vanishing value of $c_{S2}^2$ close to 0. 
The wavenumber $k$ associated with the linear regime 
of the observations of large-scale structures and weak 
lensing corresponds to $k \lesssim 300\cH_0$, where 
$\cH_0$ is the today's value of $\cH$ (at the redshift 
$z=0$). If the today's sound speed satisfies 
the condition 
\be
c_{S2}^2(z=0) \ll 10^{-5}\,,
\label{csz=0}
\ee
then the analytic solution derived above loses its validity 
for the perturbations relevant to the growth of 
large-scale structures and weak lensing.
In such cases the condition $c_{S2}^2k^2 \ll \cH^2$ 
is satisfied for the modes $k \lesssim 300\cH_0$ 
in the past, so the term $c_{S2}^2k^2$ is not the 
dominant contribution to $\beta_2$ in Eq.~(\ref{betai}).
If the masses $m_i$ are of the order $H_0$, 
the term $\beta_2 E$ as well as $E''$ and $\beta_1 E'$
would give rise to contributions at most of the 
order ${\cal H}^2 E$. 
Writing the sum of the first three terms on the l.h.s. of  Eq.~(\ref{Eeq}) as $E''+\beta_1 E'+\beta_2 E=s{\cal H}^2 E$, 
where $s$ is a time-dependent dimensionless 
factor, it follows that 
\be
s{\cal H}^2 E+\beta_4 \delta \simeq 0\,,
\ee
for the modes $k^2 \gg \cH^2$.
Taking the leading-order term in $\beta_4$ and ignoring 
the last term on the r.h.s. of Eq.~(\ref{deleqmo}), 
we can write Eq.~(\ref{deleqmo}) in the form 
(\ref{deltaeq0}) with the effective gravitational coupling 
\be
G_{\rm eff}=G \left[ 1+\frac{2 \{
(2\cH^2-2\cH'-a^2m_1^2)M_{\rm pl}^2-a^2 \bar{\rho}_M \}
(m_{2+}^2-3m_3^2+m_4^2)(m_4^2-m_T^2)}
{2M_{\rm pl}^2 \{m_1^2 (\cH^2-\cH')-2a^2m_4^2 
(m_1^2-m_4^2) \}-a^2m_1^2 \bar{\rho}_M} 
\frac{a^2}{s\cH^2} \right]\,.
\label{Geff2}
\ee
The second term in the squared bracket of Eq.~(\ref{Geff2})
does not vanish in the small-scale limit.
Provided that $m_i^2={\cal O}(m^2)$, the second 
term in the square bracket of Eq.~(\ref{Geff2}) 
is of the order $m^2/(sH^2)$. This correction can be important 
in the late Universe for $m \sim H_0$.
To compute $G_{\rm eff}$ precisely we need to 
know how the parameter $s$ varies in time, whose 
property depends on the models under consideration.
It will be of interest to study the evolution of perturbations 
for concrete models satisfying the condition (\ref{csz=0}).

\section{Conclusions}
\label{concludesec}

In this paper, we studied the cosmology in general theories of 
Lorentz-violating massive gravity by taking into 
account a perfect fluid in form of the Schutz-Sorkin 
action (\ref{SM}). This is for the purposes of
studying the dynamics of late-time cosmic acceleration 
arising from the graviton potential in the presence of matter 
and discussing observational signatures relevant to 
the growth of matter density perturbations.
The general $SO(3)$-invariant massive gravity theories that 
propagate five DOFs have the graviton potential of the 
form (\ref{Vdef}), which are parametrized by the two 
functions ${\cal U}$ and ${\cal E}$. 
By choosing the unitary gauge, we have taken into account 
the dependence of the fiducial metric 
$\delta_{ij}$ and the temporal scalar $\phi$ of the form 
$f_{ij}=b^2(\phi) \delta_{ij}$ in the 
functions ${\cal U}$ and ${\cal E}$. 

The equations of motion on the flat FLRW background are 
expressed as Eqs.~(\ref{back1}) and (\ref{back2}), where
the energy density $\rho_{\rm mg}$ and the pressure 
$P_{\rm mg}$ associated with the massive graviton 
are given by Eq.~(\ref{rhoPmg}).
Since there is the relation (\ref{HcE}) between the functions 
${\cal E}$ and ${\cal U}$, the dark energy equation of 
state arising from the graviton potential is of the form 
(\ref{wde}). The deviation of $w_{\rm DE}$ from $-1$ 
occurs for the theories in which the two conditions 
$r_b \neq 1$ and ${\rm d}{\cal U} \neq 0$ are satisfied.
While the previous studies mostly focused on the cases 
$r_b=0$ \cite{Comelli3} and $r_b=1$ \cite{Lan}, we 
extended the analysis to the theories with more general values 
of $r_b$. This amounts to considering different expansion 
rates between the scale factor $a(t)$ and the other 
time-dependent factor $b(t)$ in $f_{ij}$.

For the concrete model in which the functions ${\cal U}$ 
and ${\cal E}$ are given by Eqs.~(\ref{calU}) and (\ref{calE}), respectively, we studied 
the background cosmology relevant to the late-time 
cosmic acceleration in the presence of nonrelativistic matter and radiation.
For $r_b=1$ the ratio $Y=b/a$ remains constant, in which 
case the background cosmological dynamics is equivalent to that in the $\Lambda$CDM model.
If $r_b$ is a constant different from 1 the ratio $Y$ varies 
as $Y \propto a^{r_b-1}$, so the the relation (\ref{backcon}) 
places the constraints (\ref{v0con}) among the coefficients 
in ${\cal U}$ and ${\cal E}$.
For $r_b<1$, we have the dynamical dark energy scenario in which the equation of state changes 
in the region $w_{\rm DE}>-1$. 
On the other hand, if $r_b>1$, the phantom equation 
of state ($w_{\rm DE}<-1$) can be realized.

To discuss the stability of solutions against 
linear cosmological perturbations, we derived general conditions for the absence of ghosts and Laplacian 
instabilities of tensor and vector perturbations 
in Sec.\,\ref{tenvecsec}. 
We showed that the presence of a perfect fluid does not 
substantially modify the stabilities of tensor 
and vector modes. In our $SO(3)$-invariant massive 
gravity theories the tensor propagation speed $c_T$ is equivalent to the speed of light $c$ with the tiny graviton mass of order $H_0 \sim 10^{-33}$~eV,  so they safely 
evade the bounds recently constrained 
by LIGO \cite{LIGO,LIGO2}.
Under the conditions (\ref{veccon}) there are neither 
ghosts nor Laplacian instabilities 
for vector perturbations.

In Sec.\,\ref{scasec} we derived the second-order action of scalar perturbations in the presence of matter perturbations 
with the density contrast $\delta$. By construction the 
$SO(3)$-invariant massive gravity with the graviton potential (\ref{Vdef}) satisfies the condition $m_0^2=0$, 
under which there are two dynamical fields $E$ and $\delta$.
The conditions for the absence of ghosts and Laplacian 
instabilities associated with the field $E$ are given, 
respectively, by Eqs.~(\ref{ghostsca}) and (\ref{cscon}).
The deviation from $w_{\rm DE}=-1$, which is weighed 
by the factor $2(1-r_b){\rm d}{\cal U}/\bar{\cal U}$, 
affects the stability of scalar perturbations. 
We applied general stability conditions to the concrete 
model given by Eqs.~(\ref{calU})-(\ref{calE}) and 
showed that there are theoretically consistent parameter spaces 
in which all the stability conditions of vector and 
scalar perturbations are satisfied.

In Sec.\,\ref{growthsec} we also studied the evolution of 
the gauge-invariant nonrelativistic matter density contrast 
$\delta$ and gravitational potentials $\Psi$ and $\Phi$
to confront our massive gravity theories 
with the observations of large-scale structures 
and weak lensing. 
For the perturbations satisfying the condition (\ref{subho}), we derived the effective gravitational coupling $G_{\rm eff}$ and the gravitational slip parameter $\eta_s=-\Phi/\Psi$ by using the quasi-static approximation. As we see in 
Eqs.~(\ref{Geffqu}) and (\ref{etaqu}), 
the corrections induced by the 
massive graviton potential to their GR values 
($G_{\rm eff}=G$ and $\eta_s=1$) are suppressed 
by factors of the order $m^2a^2/(c_{S2}^2k^2)$ 
for the modes deep inside the sound horizon. 
There is a specific case $m_2^2=m_3^2$, under which 
the sound speed squared $c_{S2}^2$ can be much smaller
than 1 for $w_{\rm DE}$ close to $-1$. 
In this case the massive graviton gives rise to a correction 
to the gravitational constant without the suppression in 
the small-scale limit, see Eq.~(\ref{Geff2}).

We thus showed that the $SO(3)$-invariant theory of 
massive gravity offers an interesting possibility for giving rise to cosmological solutions with a variety of the dark 
energy equation of state, while satisfying 
the stability conditions of tensor, vector, and scalar perturbations.
It will be of interest to place observational constraints on 
concrete models like Eqs.~(\ref{calU})-(\ref{calE}) by 
using observational data associated with the cosmic expansion and growth histories.

\section*{Acknowledgements}

We thank Antonio De Felice for useful discussions.
L.\,H. thanks financial support from Dr.~Max R\"ossler, 
the Walter Haefner Foundation and the ETH Zurich
Foundation. 
S.\,T. is supported by the Grant-in-Aid for Scientific Research Fund of the JSPS No.\,16K05359 and the MEXT KAKENHI Grant-in-Aid for Scientific Research on Innovative Areas
``Cosmic Acceleration'' 
(No.\,15H05890).



\begin{thebibliography}{99}

\bibitem{rev1}
E.~J.~Copeland, M.~Sami and S.~Tsujikawa,
Int.\ J.\ Mod.\ Phys.\ D {\bf 15}, 1753 (2006)
[hep-th/0603057].

\bibitem{rev2}
T.~P.~Sotiriou and V.~Faraoni,
Rev.\ Mod.\ Phys.\  {\bf 82}, 451 (2010)
[arXiv:0805.1726 [gr-qc]].

\bibitem{rev3}
A.~De Felice and S.~Tsujikawa,
Living Rev.\ Rel.\  {\bf 13}, 3 (2010)
[arXiv:1002.4928 [gr-qc]].

\bibitem{rev4}
T.~Clifton, P.~G.~Ferreira, A.~Padilla and C.~Skordis,
Phys.\ Rept.\  {\bf 513}, 1 (2012)
[arXiv:1106.2476 [astro-ph.CO]].

\bibitem{rev5}
L.~Amendola {\it et al.} [Euclid Theory Working Group],
Living Rev.\ Rel.\  {\bf 16}, 6 (2013)
[arXiv:1206.1225 [astro-ph.CO]].
  
\bibitem{rev6} 
A.~Joyce, B.~Jain, J.~Khoury and M.~Trodden,
Phys.\ Rept.\  {\bf 568}, 1 (2015)
[arXiv:1407.0059 [astro-ph.CO]].

\bibitem{rev7}
P.~Bull {\it et al.},
Phys.\ Dark Univ.\  {\bf 12}, 56 (2016)
[arXiv:1512.05356 [astro-ph.CO]].

\bibitem{rev8}
  L.~Amendola {\it et al.} [Euclid Theory Working Group],
  Living Rev.\ Rel.\  {\bf 16}, 6 (2013)
  [arXiv:1206.1225 [astro-ph.CO]];
 L.~Amendola {\it et al.},
  arXiv:1606.00180 [astro-ph.CO].
  


\bibitem{SN1}
A.~G.~Riess \textit{et al.},
Astron.\ J.\  {\bf 116}, 1009 (1998) 
[astro-ph/9805201].

\bibitem{SN2}
S.~Perlmutter \textit{et al.},
Astrophys.\ J.\  {\bf 517}, 565 (1999) 
[astro-ph/9812133].

\bibitem{SN3}
M.~Betoule {\it et al.},
Astron.\ Astrophys.\  {\bf 568}, A22 (2014)
[arXiv:1401.4064 [astro-ph.CO]].

\bibitem{WMAP}
D.~N.~Spergel {\it et al.},
Astrophys.\ J.\ Suppl.\  {\bf 148}, 175 (2003)
[astro-ph/0302209].

\bibitem{Planck}
P.~A.~R.~Ade {\it et al.},
Astron.\ Astrophys.\  {\bf 571}, A16 (2014)
[arXiv:1303.5076 [astro-ph.CO]].

\bibitem{BAO}
D.~J.~Eisenstein {\it et al.},
Astrophys.\ J.\  {\bf 633}, 560 (2005)
[astro-ph/0501171].

\bibitem{Horndeski} 
G.~W.~Horndeski,
Int.\ J.\ Theor.\ Phys.\  {\bf 10}, 363 (1974).

\bibitem{fR1} 
T.~V.~Ruzmaikina and A.~A.~Ruzmaikin, 
Zh. \ Eksp.\ Teor.\ Fiz.\ {\bf 57}, 680 (1969) 
[Sov. Phys. - JETP 30, 372 (1970)].

\bibitem{fR2} 
A.~A.~Starobinsky,
Phys.\ Lett.\  {\bf 91B}, 99 (1980).

\bibitem{Ga1}
A.~Nicolis, R.~Rattazzi and E.~Trincherini,
Phys.\ Rev.\ D {\bf 79}, 064036 (2009)
[arXiv:0811.2197 [hep-th]].

\bibitem{Ga2} 
C.~Deffayet, G.~Esposito-Farese and A.~Vikman,
Phys.\ Rev.\ D {\bf 79}, 084003 (2009)
[arXiv:0901.1314 [hep-th]].

\bibitem{fRcos1} 
W.~Hu and I.~Sawicki,
Phys.\ Rev.\ D {\bf 76}, 064004 (2007)
[arXiv:0705.1158 [astro-ph]].

\bibitem{fRcos2} 
A.~A.~Starobinsky,
JETP Lett.\  {\bf 86}, 157 (2007)
[arXiv:0706.2041 [astro-ph]].

\bibitem{fRcos3} 
S.~A.~Appleby and R.~A.~Battye,
Phys.\ Lett.\ B {\bf 654}, 7 (2007)
[arXiv:0705.3199 [astro-ph]].

\bibitem{fRcos4} 
S.~Tsujikawa,
Phys.\ Rev.\ D {\bf 77}, 023507 (2008)
[arXiv:0709.1391 [astro-ph]].

\bibitem{Gacos1} 
R.~Gannouji and M.~Sami,
Phys.\ Rev.\ D {\bf 82}, 024011 (2010)
[arXiv:1004.2808 [gr-qc]].

\bibitem{Gacos2} 
A.~De Felice and S.~Tsujikawa,
Phys.\ Rev.\ Lett.\  {\bf 105}, 111301 (2010)
[arXiv:1007.2700 [astro-ph.CO]].

\bibitem{Heisenberg}
L.~Heisenberg,
JCAP {\bf 1405}, 015 (2014)
[arXiv:1402.7026 [hep-th]].

\bibitem{Tasinato} 
G.~Tasinato,
JHEP {\bf 1404}, 067 (2014)
[arXiv:1402.6450 [hep-th]].

\bibitem{Allys}
E.~Allys, P.~Peter and Y.~Rodriguez,
JCAP {\bf 1602}, 004 (2016)
[arXiv:1511.03101 [hep-th]].

\bibitem{Jimenez}
J.~Beltran Jimenez and L.~Heisenberg,
Phys.\ Lett.\ B {\bf 757}, 405 (2016)
[arXiv:1602.03410 [hep-th]].

\bibitem{Procacosmo}
A.~De Felice, L.~Heisenberg, R.~Kase, S.~Mukohyama, 
S.~Tsujikawa and Y.~l.~Zhang,
JCAP {\bf 1606}, 048 (2016)
[arXiv:1603.05806 [gr-qc]].

\bibitem{Procacosmo2} 
A.~De Felice, L.~Heisenberg, R.~Kase, S.~Mukohyama, S.~Tsujikawa and Y.~l.~Zhang,
Phys.\ Rev.\ D {\bf 94}, 044024 (2016)
[arXiv:1605.05066 [gr-qc]].

\bibitem{screening1} 
A.~De Felice, L.~Heisenberg, R.~Kase, S.~Tsujikawa, Y.~l.~Zhang and G.~B.~Zhao,
Phys.\ Rev.\ D {\bf 93}, 104016 (2016)
[arXiv:1602.00371 [gr-qc]].

\bibitem{screening2} 
S.~Nakamura, R.~Kase and S.~Tsujikawa,
Phys.\ Rev.\ D {\bf 96}, 084005 (2017)
[arXiv:1707.09194 [gr-qc]].

\bibitem{mareview} 
C.~de Rham,
Living Rev.\ Rel.\  {\bf 17}, 7 (2014)
[arXiv:1401.4173 [hep-th]].

\bibitem{FP} 
M.~Fierz and W.~Pauli,
Proc.\ Roy.\ Soc.\ Lond.\ A {\bf 173}, 211 (1939).

\bibitem{dis1} 
H.~van Dam and M.~J.~G.~Veltman,
Nucl.\ Phys.\ B {\bf 22}, 397 (1970).

\bibitem{dis2} 
V.~I.~Zakharov,
JETP Lett.\  {\bf 12}, 312 (1970)
[Pisma Zh.\ Eksp.\ Teor.\ Fiz.\  {\bf 12}, 447 (1970)].

\bibitem{Vainshtein} 
A.~I.~Vainshtein,
Phys.\ Lett.\  {\bf 39B}, 393 (1972).

\bibitem{BD} 
D.~G.~Boulware and S.~Deser,
Phys.\ Rev.\ D {\bf 6}, 3368 (1972).

\bibitem{dRGT} 
C.~de Rham, G.~Gabadadze and A.~J.~Tolley,
Phys.\ Rev.\ Lett.\  {\bf 106}, 231101 (2011)
[arXiv:1011.1232 [hep-th]].

\bibitem{deRham10} 
C.~de Rham and G.~Gabadadze,
Phys.\ Rev.\ D {\bf 82}, 044020 (2010)
[arXiv:1007.0443 [hep-th]].

\bibitem{Hassan} 
S.~F.~Hassan and R.~A.~Rosen,
Phys.\ Rev.\ Lett.\  {\bf 108}, 041101 (2012)
[arXiv:1106.3344 [hep-th]].

\bibitem{DAmico} 
G.~D'Amico, C.~de Rham, S.~Dubovsky, G.~Gabadadze, 
D.~Pirtskhalava and A.~J.~Tolley,
Phys.\ Rev.\ D {\bf 84}, 124046 (2011)
[arXiv:1108.5231 [hep-th]].

\bibitem{deRham:2010tw} 
C.~de Rham, G.~Gabadadze, L.~Heisenberg and D.~Pirtskhalava,
Phys.\ Rev.\ D {\bf 83}, 103516 (2011)
[arXiv:1010.1780 [hep-th]].

\bibitem{deRham:2011by} 
C.~de Rham and L.~Heisenberg,
Phys.\ Rev.\ D {\bf 84}, 043503 (2011)
[arXiv:1106.3312 [hep-th]];
  L.~Heisenberg, R.~Kimura and K.~Yamamoto,
  Phys.\ Rev.\ D {\bf 89}, 103008 (2014)
  [arXiv:1403.2049 [hep-th]].

\bibitem{Burrage:2011cr} 
  C.~Burrage, C.~de Rham, L.~Heisenberg and A.~J.~Tolley,
  JCAP {\bf 1207}, 004 (2012)
  [arXiv:1111.5549 [hep-th]];
  C.~de Rham, G.~Gabadadze, L.~Heisenberg and D.~Pirtskhalava,
  Phys.\ Rev.\ D {\bf 87}, no. 8, 085017 (2013)
  [arXiv:1212.4128 [hep-th]].
 
\bibitem{Chamseddine:2011bu} 
  A.~H.~Chamseddine and M.~S.~Volkov,
  Phys.\ Lett.\ B {\bf 704}, 652 (2011)
  [arXiv:1107.5504 [hep-th]].

\bibitem{Koyama:2011xz} 
K.~Koyama, G.~Niz and G.~Tasinato,
Phys.\ Rev.\ Lett.\  {\bf 107}, 131101 (2011)
[arXiv:1103.4708 [hep-th]].
 
\bibitem{Gratia:2012wt} 
P.~Gratia, W.~Hu and M.~Wyman,
Phys.\ Rev.\ D {\bf 86}, 061504 (2012)
[arXiv:1205.4241 [hep-th]].

\bibitem{Comelli:2011zm} 
D.~Comelli, M.~Crisostomi, F.~Nesti and L.~Pilo,
JHEP {\bf 1203}, 067 (2012)
[arXiv:1111.1983 [hep-th]].
 
\bibitem{Comelli:2012db} 
D.~Comelli, M.~Crisostomi and L.~Pilo,
JHEP {\bf 1206}, 085 (2012)
[arXiv:1202.1986 [hep-th]].

\bibitem{Volkov:2012zb} 
M.~S.~Volkov,
Phys.\ Rev.\ D {\bf 86}, 104022 (2012)
[arXiv:1207.3723 [hep-th]].

\bibitem{Emir} 
A.~E.~Gumrukcuoglu, C.~Lin and S.~Mukohyama,
JCAP {\bf 1203}, 006 (2012)
[arXiv:1111.4107 [hep-th]].

\bibitem{Tolley} 
M.~Fasiello and A.~J.~Tolley,
JCAP {\bf 1211}, 035 (2012)
[arXiv:1206.3852 [hep-th]].

\bibitem{Naruko} 
D.~Langlois and A.~Naruko,
Class.\ Quant.\ Grav.\  {\bf 29}, 202001 (2012)
[arXiv:1206.6810 [hep-th]].

\bibitem{DGM} 
A.~De Felice, A.~E.~Gumrukcuoglu and S.~Mukohyama,
Phys.\ Rev.\ Lett.\  {\bf 109}, 171101 (2012)
[arXiv:1206.2080 [hep-th]].

\bibitem{maco1} 
C.~de Rham, L.~Heisenberg and R.~H.~Ribeiro,
Class.\ Quant.\ Grav.\  {\bf 32}, 035022 (2015)
[arXiv:1408.1678 [hep-th]].

\bibitem{maco2} 
C.~de Rham, L.~Heisenberg and R.~H.~Ribeiro,
Phys.\ Rev.\ D {\bf 90}, 124042 (2014)
[arXiv:1409.3834 [hep-th]].

\bibitem{Heisenberg:2014rka} 
L.~Heisenberg,
Class.\ Quant.\ Grav.\  {\bf 32}, no. 10, 105011 (2015)
[arXiv:1410.4239 [hep-th]].

\bibitem{Gumrukcuoglu:2014xba} 
A.~Emir Gumrukcuoglu, L.~Heisenberg and S.~Mukohyama,
JCAP {\bf 1502}, 022 (2015)
[arXiv:1409.7260 [hep-th]].
  
\bibitem{Heisenberg:2016spl} 
L.~Heisenberg and A.~Refregier,
JCAP {\bf 1609}, 020 (2016)
[arXiv:1604.07306 [gr-qc]];
  L.~Heisenberg and A.~Refregier,
  Phys.\ Lett.\ B {\bf 762}, 131 (2016)
  [arXiv:1604.07680 [astro-ph.CO]].

\bibitem{Higuchi} 
A.~Higuchi,
Nucl.\ Phys.\ B {\bf 282}, 397 (1987).

\bibitem{Rubakov} 
V.~A.~Rubakov,
hep-th/0407104.

\bibitem{Dub1} 
S.~L.~Dubovsky,
JHEP {\bf 0410}, 076 (2004)
[hep-th/0409124].

\bibitem{Dub2} 
S.~L.~Dubovsky, P.~G.~Tinyakov and I.~I.~Tkachev,
Phys.\ Rev.\ Lett.\  {\bf 94}, 181102 (2005)
[hep-th/0411158].

\bibitem{Dub3} 
S.~L.~Dubovsky, P.~G.~Tinyakov and I.~I.~Tkachev,
Phys.\ Rev.\ D {\bf 72}, 084011 (2005)
[hep-th/0504067].

\bibitem{Beb} 
M.~V.~Bebronne and P.~G.~Tinyakov,
Phys.\ Rev.\ D {\bf 76}, 084011 (2007)
[arXiv:0705.1301 [astro-ph]].

\bibitem{Gaba} 
G.~Gabadadze and L.~Grisa,
Phys.\ Lett.\ B {\bf 617}, 124 (2005)
[hep-th/0412332].

\bibitem{Blas} 
D.~Blas, D.~Comelli, F.~Nesti and L.~Pilo,
Phys.\ Rev.\ D {\bf 80}, 044025 (2009)
[arXiv:0905.1699 [hep-th]].

\bibitem{Lin1} 
C.~Lin and M.~Sasaki,
Phys.\ Lett.\ B {\bf 752}, 84 (2016)
[arXiv:1504.01373 [astro-ph.CO]].

\bibitem{Lin2} 
G.~Domenech, T.~Hiramatsu, C.~Lin, M.~Sasaki, M.~Shiraishi and Y.~Wang,
JCAP {\bf 1705}, 034 (2017)
[arXiv:1701.05554 [astro-ph.CO]].

\bibitem{Lin3} 
S.~Kuroyanagi, C.~Lin, M.~Sasaki and S.~Tsujikawa,
arXiv:1710.06789 [gr-qc].

\bibitem{Comelli1} 
D.~Comelli, F.~Nesti and L.~Pilo,
Phys.\ Rev.\ D {\bf 87}, 124021 (2013)
[arXiv:1302.4447 [hep-th]].

\bibitem{Comelli2} 
D.~Comelli, F.~Nesti and L.~Pilo,
JHEP {\bf 1307}, 161 (2013)
[arXiv:1305.0236 [hep-th]].

\bibitem{Comelli3} 
D.~Comelli, F.~Nesti and L.~Pilo,
JCAP {\bf 1405}, 036 (2014)
[arXiv:1307.8329 [hep-th]].

\bibitem{Lan} 
D.~Langlois, S.~Mukohyama, R.~Namba and A.~Naruko,
Class.\ Quant.\ Grav.\  {\bf 31}, 175003 (2014)
[arXiv:1405.0358 [hep-th]].

\bibitem{Sorkin}
B.~F.~Schutz and R.~Sorkin,
Annals Phys.\  {\bf 107}, 1 (1977).

\bibitem{DGS}
A.~De Felice, J.~M.~Gerard and T.~Suyama,
Phys.\ Rev.\ D {\bf 81}, 063527 (2010)
[arXiv:0908.3439 [gr-qc]].

\bibitem{LIGO} 
B.~P.~Abbott {\it et al.}, 
Astrophys.\ J.\  {\bf 848}, no. 2, L13 (2017).

\bibitem{LIGO2} 
B.~P.~Abbott {\it et al.}, 
Phys.\ Rev.\ Lett.\  {\bf 118}, no. 22, 221101 (2017)
[arXiv:1706.01812 [gr-qc]].

\bibitem{ADM} 
R.~L.~Arnowitt, S.~Deser and C.~W.~Misner,
Gen.\ Rel.\ Grav.\  {\bf 40}, 1997 (2008)
[gr-qc/0405109].

\bibitem{mini1} 
A.~De Felice and S.~Mukohyama,
Phys.\ Lett.\ B {\bf 752}, 302 (2016)
[arXiv:1506.01594 [hep-th]].

\bibitem{mini2} 
A.~De Felice and S.~Mukohyama,
JCAP {\bf 1604}, 028 (2016)
[arXiv:1512.04008 [hep-th]].

\bibitem{nl1} 
M.~Jaccard, M.~Maggiore and E.~Mitsou,
Phys.\ Rev.\ D {\bf 88}, 044033 (2013)
[arXiv:1305.3034 [hep-th]].

\bibitem{nl2} 
M.~Maggiore,
Phys.\ Rev.\ D {\bf 89}, 043008 (2014)
[arXiv:1307.3898 [hep-th]].

\bibitem{nl3} 
L.~Modesto and S.~Tsujikawa,
Phys.\ Lett.\ B {\bf 727}, 48 (2013)
[arXiv:1307.6968 [hep-th]].

\bibitem{Bardeen} 
J.~M.~Bardeen,
Phys.\ Rev.\ D {\bf 22}, 1882 (1980).

\bibitem{Uzan} 
C.~Schimd, J.~P.~Uzan and A.~Riazuelo,
Phys.\ Rev.\ D {\bf 71}, 083512 (2005)
[astro-ph/0412120].

\bibitem{Sapone} 
L.~Amendola, M.~Kunz and D.~Sapone,
JCAP {\bf 0804}, 013 (2008)
[arXiv:0704.2421 [astro-ph]].

\bibitem{quasi1} 
B.~Boisseau, G.~Esposito-Farese, D.~Polarski and A.~A.~Starobinsky,
Phys.\ Rev.\ Lett.\  {\bf 85}, 2236 (2000)
[gr-qc/0001066].

\bibitem{quasi2}
S.~Tsujikawa,
Phys.\ Rev.\ D {\bf 76}, 023514 (2007)
[arXiv:0705.1032 [astro-ph]].

\bibitem{quasi3}
A.~De Felice, T.~Kobayashi and S.~Tsujikawa,
Phys.\ Lett.\ B {\bf 706}, 123 (2011)
[arXiv:1108.4242 [gr-qc]].


\end{thebibliography}
\end{document}